\documentclass[a4paper,12pt]{article}

\pdfoutput=1

\usepackage{mathptmx}       % selects Times Roman as basic font
\usepackage{helvet}         % selects Helvetica as sans-serif font
\usepackage{courier}        % selects Courier as typewriter font
\usepackage{type1cm}        % activate if the above 3 fonts are
\usepackage{amsfonts} 
\usepackage{algorithm} 
\usepackage{makeidx}         % allows index generation
\usepackage{graphicx}        % standard LaTeX graphics tool
\usepackage{url}       

\usepackage{multicol}        % used for the two-column index
\usepackage[bottom]{footmisc}% places footnotes at page bottom
\usepackage{amsthm}
\usepackage{amsmath}

\floatstyle{ruled}
\newfloat{algorithm}{tbp}{loa}
\providecommand{\algorithmname}{Algorithm}
\floatname{algorithm}{\protect\algorithmname}

\numberwithin{equation}{section}
\numberwithin{figure}{section}
\theoremstyle{plain}
\newtheorem{thm}{\protect\theoremname}[section]
\theoremstyle{definition}
\newtheorem{defn}[thm]{\protect\definitionname}
\theoremstyle{remark}

\theoremstyle{plain}

\theoremstyle{remark}

\theoremstyle{plain}
\newtheorem{corollary}[thm]{\protect\corollaryname}
\theoremstyle{plain}
\newtheorem{proposition}[thm]{\protect\propositionname}

\providecommand{\claimname}{Claim}
\providecommand{\definitionname}{Definition}
\providecommand{\lemmaname}{Lemma}
\providecommand{\remarkname}{Remark}
\providecommand{\theoremname}{Theorem}
\providecommand{\corollaryname}{Corollary}
\providecommand{\propositionname}{Proposition}

\usepackage{authblk}

\makeindex             % used for the subject index
\title{Fourier Phase Retrieval:\\  Uniqueness and Algorithms}

\author[1]{Tamir Bendory\thanks{tamir.bendory@princeton.edu}}
\author[2]{Robert Beinert\thanks{robert.beinert@uni-graz.at. The Institute of Mathematics and Scientific Computing
		 is a member of NAWI Graz (http://www.nawigraz.at).  The author
		 is supported by the Austrian Science Fund (FWF) within the project P~28858.}}
\author[3]{Yonina C. Eldar\thanks{yonina@ee.technion.ac.il. The author is supported by the European Union’s Horizon
		2020 research and innovation program under grant agreement no. 646804-ERC-COG-BNYQ, and from the Israel Science Foundation under Grant no. 335/14.}}

\affil[1]{The Program in Applied and Computational Mathematics,
			Princeton University, Princeton, NJ, USA}
\affil[2]{Institute of Mathematics
	 and Scientific Computing, University of Graz, Heinrichstra\ss e 36, 8010
	 Graz, Austria}
\affil[3]{The Andrew and Erna Viterbi Faculty of Electrical
	Engineering, Technion - Israel Institute of Technology, Haifa, Israel}

\date{}

\begin{document}

%\title*{Fourier Phase Retrieval: Uniqueness and Algorithms}
%  \titlerunning{Fourier Phase Retrieval: Uniqueness and Algorithms}
\maketitle

%
%\author{$^{\dagger}$ \thanks{The Program in Applied and Computational Mathematics,
%		Princeton University, Princeton, NJ, USA
%		, \url{tamir.bendory@princeton.edu}},}
%
%\author{Tamir Bendory, Robert Beinert and Yonina C. Eldar}
%% Use \authorrunning{Short Title} for an abbreviated version of
%% your contribution title if the original one is too long
%\institute{Tamir Bendory \at 
% \and Robert Beinert \at Institute of Mathematics
% and Scientific Computing, University of Graz, Heinrichstra\ss e 36, 8010
% Graz, Austria, \email{robert.beinert@uni-graz.at}. The Institute of Mathematics and Scientific Computing
% is a member of NAWI Graz (\url{http://www.nawigraz.at/}).  The author
% is supported by the Austrian Science Fund (FWF) within the project P~28858.
%\and Yonina C. Eldar \at The Andrew and Erna Viterbi Faculty of Electrical
%Engineering, Technion - Israel Institute of Technology, Haifa, Israel,  \email{yonina@ee.technion.ac.il}. The author is supported by the European Union’s Horizon
%2020 research and innovation program under grant agreement no. 646804-ERC-COG-BNYQ, and from the Israel Science Foundation under Grant no. 335/14.}
%%
% Use the package "url.sty" to avoid
% problems with special characters
% used in your e-mail or web address
%

\abstract{ The problem of recovering a signal from its phaseless Fourier transform measurements, called Fourier phase retrieval, arises in many applications in engineering and science. Fourier phase retrieval  poses fundamental theoretical and algorithmic challenges. In general, there is no unique mapping between a one-dimensional signal and its Fourier magnitude and therefore the problem is ill-posed. Additionally, while almost  all multidimensional signals are uniquely mapped to their Fourier magnitude, the performance of existing algorithms is generally not well-understood.  In this chapter we survey methods to guarantee uniqueness in Fourier phase retrieval. We then present different algorithmic approaches to retrieve the signal in practice. We conclude  by outlining some of the main open questions in this field.  
 }

%~~~~~~~~~~~~~~~~~~~~~~~~~~~~~~~~~~~~~~~~~~~~~~~~~~~~~~~~~~~~~%

%
%\begin{acknowledgement}
%The second author gratefully acknowledges the funding of this work by the Austrian Science Fund (FWF) within the project P 28858.  The Institute of Mathematics and Scientific Computing of the University of Graz, with which the second author is affiliated, is a member of NAWI Graz (http://www.nawigraz.at/).
%\end{acknowledgement}

%~~~~~~~~~~~~~~~~~~~~~~~~~~~~~~~~~~~~~~~~~~~~~~~~~~~~~~~~~~~~~%

\section{Introduction} \label{sec:intro}

The task of recovering a signal from its Fourier transform magnitude,
called \emph{Fourier phase retrieval}, arises in many areas in
engineering and science. The problem has a rich history, tracing back
to 1952 \cite{sayre1952some}. Important examples for Fourier phase
retrieval naturally appear in many optical settings since optical
sensors, such as a charge-coupled device (CCD) and the human eye, are
insensitive to phase information of the light wave. A typical example
is coherent diffraction imaging (CDI) which is used in a variety of
imaging techniques \cite{bertolotti2012non, chapman2006femtosecond,
  chen2009multiple, miao1999extending, robinson2001reconstruction,
  sandberg2008high}.  In CDI, an object is illuminated with a coherent
electro-magnetic wave and the far-field intensity diffraction pattern
is measured. This pattern is proportional to the object's Fourier
transform and therefore the measured data is proportional to its
Fourier magnitude.  Phase retrieval also played a key role in the
development of the DNA double helix model
\cite{garwin2010century}. This discovery awarded Watson, Crick and
Wilkins the Nobel prize in Physiology or Medicine in 1962
\cite{nobelprize}.  Additional examples for applications in which
Fourier phase retrieval appear are X-ray crystallography, speech
recognition, blind channel estimation, astronomy, computational
biology, alignment and blind deconvolution
\cite{baykal2004blind,bendory2017bispectrum, fienup1987phase,
  harrison1993phase, rabiner1993fundamentals, shechtman2015phase,
  stefik1978inferring, walther1963question, yeh2015experimental}.

Fourier phase retrieval has been a long-standing problem since it
raises difficult challenges. In general, there is no unique mapping
between a one-dimensional signal and its Fourier magnitude and
therefore the problem is ill-posed. Additionally, while almost all
multidimensional signals are uniquely mapped to their Fourier
magnitude, the performance and stability of existing algorithms is
generally not well-understood. In particular, it is not clear when
given methods recover the true underlying signal.  To simplify the
mathematical analysis, in recent years attention has been devoted to a
family of related problems, frequently called \emph{generalized phase
  retrieval}. This refers to the setting in which the measurements are
the phaseless inner products of the signal with known vectors.
Particularly, the majority of works studied inner products with random
vectors.  Based on probabilistic considerations, a variety of convex
and non-convex algorithms were suggested, equipped with stability
guarantees from near-optimal number of measurements; see
\cite{balan2006signal, bandeira2014saving, candes2015phase,
  chen2015solving, eldar2014phase, goldstein2016phasemax,
  soltanolkotabi2017structured,waldspurger2015phase,wang2016solving} to name a few works
along these lines.

Here, we focus on the original Fourier phase retrieval problem and study it in detail. We begin by considering the ambiguities of Fourier phase retrieval \cite{Bei15,BP15,BP16}. We show that while in general a one-dimensional signal cannot be determined from its Fourier magnitude, there are several exceptional cases, such as minimum phase \cite{haung2016} and sparse signals \cite{jaganathan2013sparse,RCLV13}. For general signals, one can guarantee uniqueness by taking multiple measurements, each one with a different mask. This setup is called masked phase retrieval \cite{candes2015phase,jaganathan2015phase_mask} and has several interesting special cases, such as the short-time Fourier transform (STFT) phase retrieval \cite{bendory2016non,eldar2015sparse,jaganathan2016stft} and vectorial phase retrieval \cite{leshem2016direct,RDN13,raz2014direct,raz2011vectorial}. For all aforementioned setups, we present algorithms and discuss their properties. We also study the closely--related Frequency-Resolved Optical Gating (FROG) methods \cite{bendory2017on} and  multidimensional Fourier phase retrieval~\cite{kogan20162d}.  

%We will also  consider the case of  which is an important special of masked phase retrieval.   

%The goal of this chapter is twofold. First, to provide a comprehensive
%survey of theoretical guarantees and algorithms for Fourier phase
%retrieval. Second, to stress the gaps between practitioners, existing
%algorithms and our theoretical understanding. By stressing those gaps
%we hope to motivate more researchers to focus their attention on this
%field of research. 

The outline of this chapter is as follows. In Section
\ref{sec:problem_formulation} we formulate the Fourier phase retrieval
problem. We also introduce several of its variants, such as  masked Fourier phase
retrieval and STFT phase retrieval.  In
Section \ref{sec:uniqueness} we discuss the fundamental problem of
uniqueness. Namely, conditions under which there is a unique
mapping between a signal and its phaseless measurements.  Section
\ref{sec:algorithms} is devoted to different algorithmic approaches to
recover a signal from its phaseless measurements. Section
\ref{sec:conclusion} concludes the chapter and outlines some  open questions. We hope that highlighting the gaps in the theory of phase retrieval will motivate more research on these issues.

%%% Local Variables:
%%% mode: latex
%%% TeX-master: "draft4"
%%% End:

\section{Problem formulation} \label{sec:problem_formulation}

In this section, we formulate the Fourier phase retrieval problem and
introduce notation. 

Let $x\in\mathbb{C}^N$ be the underlying signal we wish to recover. 
In Fourier phase retrieval  the measurements are given by
\begin{equation} 
  \begin{split} \label{eq:PR_classical}
    &y[k]=\left \vert \sum_{n=0}^{N-1}x[n]e^{-2\pi j kn/\tilde{N}}\right\vert^2, \quad  k=0,\dots,K-1.
  \end{split}
\end{equation}
Unless otherwise mentioned, we consider the over-sampled Fourier transform, i.e., $\tilde N = K= 2N-1$, since in this case the acquired data is equivalent to the auto-correlation of $x$ as explained in Section \ref{sec:trivial-non-trivial}.  We refer to this
case as the \emph{classical phase retrieval problem}.  As will be
discussed in the next section, in general the classical phase
retrieval problem is ill-posed.  Nevertheless, some special structures may impose uniqueness. Two
important examples are sparse signals obeying a non-periodic 
support \cite{jaganathan2013sparse,RCLV13} and minimum phase signals \cite{haung2016}; see Section~\ref{sec:ensuring-uniqueness}.

For general signals, a popular method  to guarantee a unique mapping between the signal and its phaseless
Fourier measurements is by utilizing several {masks} to introduce
redundancy in the acquired data. 
 In this case, the measurements are given by 
\begin{equation} 
  \begin{split} \label{eq:PR}
    y[m,k]=\left \vert \sum_{n=0}^{N-1}x[n]d_{m}[n]e^{-2\pi j kn/
        \tilde N}\right\vert^2, \quad
     k=0,\dots,K-1, \quad m=0,\dots,M-1, 
  \end{split}
\end{equation}
where $d_m$ are $M$ known {masks}.  In
matrix notation, this model can be written as
\begin{equation} 
  \begin{split} \label{eq:PR_vector}
    y[m,k]=\left\vert f_k^*D_mx\right\vert^2 ,\quad  k=0,\dots,K-1, \quad m=0,\dots,M-1,
  \end{split}
\end{equation}
where $f_k^*$ is the $k$th row of the DFT matrix
$F\in\mathbb{C}^{K\times N}$ and $D_m\in\mathbb{C}^{N\times N}$ is a diagonal matrix that contains the entries of the $m$th mask. 
Classical phase retrieval is a special case in which $M=1$ and $D_0=I_N$ is the identity matrix. 

\begin{figure} 
        \centering
        \includegraphics[scale=0.6]{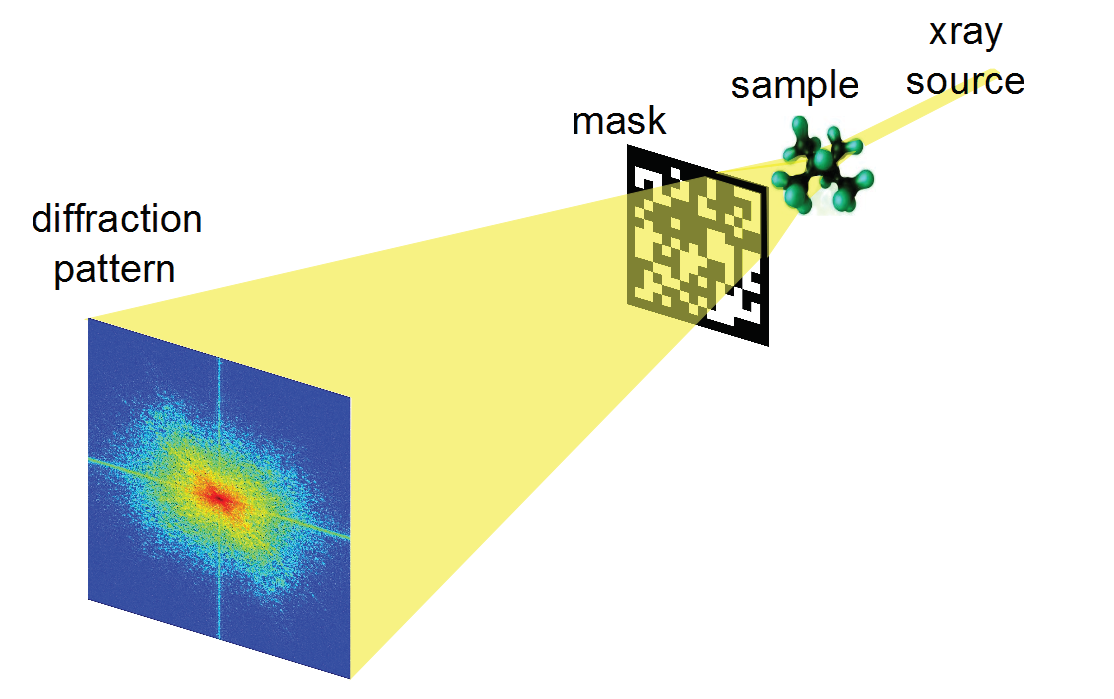}
        \caption{An illustration of a typical masked phase retrieval setup (courtesy of~\cite{candes2015phase}).}       \label{fig:maskedPR}
\end{figure}

There are several experimental
techniques to generate masked Fourier measurements in optical setups
\cite{candes2015phase}. One method is to insert a mask or a phase
plate after the object \cite{liu2008phase}. Another possibility is to
modulate the illuminating beam by an optical grating
\cite{loewen1997diffraction}. A third alternative is oblique
illumination by illuminating beams hitting the object at specified
angles \cite{faridian2010nanoscale}. An illustration of a masked phase retrieval setup is shown in Fig.~\ref{fig:maskedPR}.

An interesting special case of masked phase retrieval is signal reconstruction from
phaseless STFT measurements. Here, all masks are translations of a
reference mask, i.e.,\  $d_m[n]=d[mL-n]$, where $L$ is a parameter that
determines the overlapping factor between adjacent windows.
Explicitly, the STFT phase retrieval problem takes on the form 
\begin{equation} \label{eq:PR_STFT}
  \begin{split}
    &y[m,k]=\left \vert \sum_{n=0}^{N-1}x[n]d[mL-n]e^{-2\pi j
        kn/\tilde N}\right\vert^2  ,\\ &k=0,\dots,K-1, \quad m=0,\dots,\left\lceil N/L\right\rceil -1.
  \end{split}
\end{equation} 
The reference mask $d$ is referred to as \emph{STFT window}. We denote
the length of the STFT window by $W$, namely, $d[n]=0$ for
$n=W,\dots,N-1$ for some $W\leq N$.

\begin{figure} 
        \centering
        \includegraphics[scale=0.6]{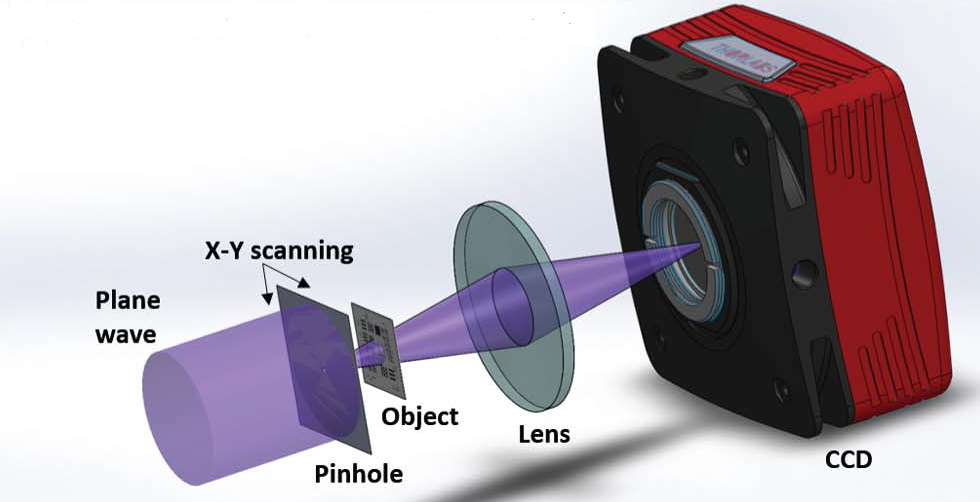}
        \caption{An illustration of a conventional ptychography setup (courtesy of~\cite{sidorenko2016single}).}        \label{fig:ptychography}
\end{figure}

The problem of recovering a signal from its STFT magnitude arises in
several applications in optics and speech processing. Particularly, it serves as the model of a
popular variant of an ultra-short laser pulse measurement technique
called Frequency-Resolved Optical Gating (FROG) which is introduced in
Section~\ref{sec:FROG} (the variant is referred to as X-FROG)
\cite{bendory2017on,trebino2012frequency}. 
Another application is ptychography in
which a moving probe (pinhole) is used to sense multiple diffraction
measurements \cite{maiden2011superresolution,
  marchesini2015alternating, rodenburg2008ptychography}.  An illustration of a conventional ptychography setup is given in Fig.~\ref{fig:ptychography}.
A closely-related problem is Fourier ptychography \cite{yeh2015experimental}.

The next section is devoted to the question of uniqueness. Namely,
under what conditions on the signal $x$ and the masks $d_m$
there exists a unique mapping between the signal and the phaseless
measurements. In Section \ref{sec:algorithms} we survey different algorithmic approaches to recover the underlying signal $x$ from the acquired data.

%%% Local Variables:
%%% mode: latex
%%% TeX-master: "draft4"
%%% End:

\section{Uniqueness guarantees} \label{sec:uniqueness} 

The aim of this section is to survey several approaches to ensure
uniqueness of the discrete phase retrieval problem.  We begin  our
study in Section~\ref{sec:trivial-non-trivial} by considering the ambiguities arising   in classical phase
retrieval and provide a complete characterization of the solution set. Although the problem is highly ambiguous in general, uniqueness
can be ensured if additional information about the signal is 
available.  In Section~\ref{sec:ensuring-uniqueness}, we first consider uniqueness guarantees based on the
knowledge of the absolute value or phase of some signal entries.
Next, we study sparse and minimum phase signals, which are
uniquely defined by their Fourier magnitude and can be recovered by
stable  algorithms.  In Sections~\ref{sec:phase-retri-mask} and~\ref{sec:phase-retri-stft} we show that for general signals, the ambiguities
 may be avoided by measuring the Fourier magnitudes of the interaction
of the true signal with multiple deterministic masks or with several
shifts of a fixed window. In Section~\ref{sec:FROG} we study uniqueness guarantees for
the closely--related FROG methods.
 Finally,  in Section~\ref{sec:multidimens-phase} we survey the multidimensional phase retrieval problems and their properties that differ significantly from the one-dimensional setting.

\subsection{Trivial and non-trivial ambiguities}
\label{sec:trivial-non-trivial}

Considering the measurement model \eqref{eq:PR_classical} of the
classical phase retrieval problem, we immediately observe that the
true signal $x \in \mathbb C^N$ cannot be recovered uniquely.  For
instance, the rotation (multiplication with a unimodular factor), the
translation, or the conjugate reflection do not modify the Fourier
magnitudes.  Without further a priori constraints, the unknown signal
$x$ is hence only determined up to these so-called \emph{trivial
ambiguities}, which are of minor interest.  Beside the trivial
ambiguities, the classical phase retrieval problem usually has a
series of further non-trivial solutions, which can strongly differ
from the true signal.  For instance, the two non-trivially different
signals
\begin{equation*}
  x_1 = (1, 0, -2, 0, -2)^T
  \qquad\text{and}\qquad
  x_2 = ((1-\sqrt 3), 0, 1, 0 , (1+\sqrt 3))^T
\end{equation*}
yield the same Fourier magnitudes $y[k]$ in~\eqref{eq:PR_classical}; 
see~\cite{shechtman2015phase}.

To characterize the occurring non-trivial ambiguities, one can exploit
the close relation between the given Fourier magnitudes $y[k]$ with
$k=0, \dots, 2N-2$ in \eqref{eq:PR_classical} and the autocorrelation
signal
\begin{equation*}
  a[n] = \sum_{m=0}^{N-1} \overline{x [m]} \, x[m+n],
  \qquad n=-N+1, \dots, N-1,
\end{equation*}
with $x[n] = 0$ for $n<0$ and $n \ge N$, \cite{BP15, BS79}.  For this
purpose, we consider the product of the polynomial
$X(z) = \sum_{n=0}^{N-1} x[n] \, z^n$ and the reversed polynomial
$\tilde X(z) = z^{N-1} \overline X (z^{-1})$, where $\overline X$
denotes the polynomial with conjugate coefficients.  Note that
$X(z^{-1})$ coincides with the usual $z$-transform of the signal $x
\in \mathbb C^N$.  Assuming
that $x[0] \ne 0$ and $x[N-1] \ne 0$, we have
\begin{equation*}
  X(z) \, \tilde X(z)
  = z^{N-1} \sum_{n=0}^{N-1} x[n] \, z^n  \cdot \sum_{m=0}^{N-1} \overline{x [m]} \,
    \, z^{-m}
    =\sum_{n=0}^{2N-2} a [n-N+1] \, z^n
    =: A(z),
\end{equation*}
where $A(z)$ is the autocorrelation polynomial  of degree $2N-2$.

Since the Fourier magnitude \eqref{eq:PR_classical} can be written as
\begin{equation*}
  y[k]
  = e^{2 \pi jk (N-1) / \tilde N} \, X\bigl(e^{-2 \pi j k / \tilde N}\bigr) \,
  \tilde X\bigl(e^{-2 \pi j k / \tilde N}\bigr)
  = e^{2 \pi j (N-1) / \tilde N} \, A\bigl(e^{-2 \pi j k / \tilde N}\bigr),
\end{equation*}
the autocorrelation polynomial $A(z)$ is completely determined by the
$2N-1$ samples $y[k]$. The classical phase retrieval problem is thus
equivalent to the recovery of $X(z)$ from
\begin{equation*}
  A(z) =  X(z) \, \tilde X(z).
\end{equation*}
Comparing the roots of $X(z)$ and $\tilde X(z)$, we observe that the
roots of the autocorrelation polynomial $A(z)$ occur in reflected
pairs
\raisebox{0pt}[0pt][0pt]{$(\gamma_j, \overline \gamma_j^{\,-1})$} with
respect to the unit circle.  The main problem in the recovery of
$X(z)$ is now to decide whether $\gamma_j$ or
$\overline \gamma_j^{\,-1}$ is a root of $X(z)$.  On the basis of this
observation, all ambiguities -- trivial and non-trivial -- are
characterized in the following way.

\begin{thm}
  \label{the:char-ambi}
  {\upshape \cite{BP15}}
  Let $x \in \mathbb C^N$ be a complex-valued signal with $x[0] \ne 0$ and
  $x[N-1] \ne 0$ and with the Fourier magnitudes $y[k]$, $k=0, \dots,
  2N-2$, in \eqref{eq:PR_classical}.  Then the
  polynomial $X'(z) = \sum_{n=0}^{N-1} x'[n] \, z^n$ of each signal $x' \in \mathbb
  C^N$ with $y'[k] = y[k]$ can be written as
  \begin{equation*}
    X' (z) 
    = e^{j\alpha} \sqrt{\vert a[N-1]
      \prod_{i=1}^{N-1} \vert \beta_i \vert^{-1}} \cdot
    \prod_{i=1}^{N-1} (z - \beta_i),
  \end{equation*}
  where $\alpha \in [-\pi, \pi)$, and where $\beta_i$ is chosen from
  the reflected zero pairs $(\gamma_i, \overline \gamma_i^{\,-1})$ of
  the autocorrelation polynomial $A(z)$.  Moreover, up to $2^{N-2}$ of
  these solutions may be non-trivially different.
\end{thm}

Since the support length $N$ of the true signal $x$ is directly
encoded in the degree of the autocorrelation polynomial, all signals
$x'$ with $y'[k]=y[k]$ in Theorem~\ref{the:char-ambi} have the same
length, and the trivial shift ambiguity does not occur.  The
multiplication by $e^{j\alpha}$ is related to the trivial rotation
ambiguity.  The trivial conjugate reflection ambiguity is also covered
by Theorem~\ref{the:char-ambi}, since this corresponds to the
reflection of all zeros $\beta_i$ at the unit circle and to an
appropriate rotation of the whole signal.  Hence, at least two of the
$2^{N-1}$ possible zero sets $\{\beta_1, \dots, \beta_{N-1}\}$ always
correspond to the same non-trivial solution, which implies that the
number of non-trivial solutions of the classical phase retrieval
problem is bounded by $2^{N-2}$.

The actual number of non-trivial ambiguities for a specific phase
retrieval problem, however, strongly depends on the zeros of the true
solution.  If $L$ denotes the number of  zero pairs
\raisebox{0pt}[0pt][0pt]{$(\gamma_\ell, \overline
  \gamma\kern0pt_\ell^{\,-1})$}
of the autocorrelation polynomial $A(z)$ not lying on the unit circle,
and $m_\ell$ the multiplicities of these zeros, then the different zero
sets $\{\beta_1, \dots, \beta_{N-1}\}$ in Theorem~\ref{the:char-ambi}
can consist of $s_\ell$ roots $\gamma_\ell$ and $(m_\ell - s_\ell)$
roots $\overline \gamma_\ell^{\,-1}$, where $s_\ell$ is an integer
between $0$ and $m_\ell$.  Due to the trivial conjugation and
reflection ambiguity, the corresponding phase retrieval problem has
exactly
\begin{equation*}
  \left\lceil \frac{1}{2} \prod_{\ell=1}^L (m_\ell + 1) \right\rceil
\end{equation*}
 non-trivial solutions
\cite{Bei15, Fej16}.  If, for instance, all zero pairs
$(\gamma_\ell, \overline \gamma\kern0pt_\ell^{\,-1})$ are unimodular,
then the problem is even uniquely solvable.

% Using the characterization in Theorem~\ref{the:char-ambi}, one can
% further show that every solution of the classical phase retrieval
% problem can be represented by a factorization of the true signal with
% respect to the convolution, and by rotations, shifts, and conjugate
% reflections of the appearing factors \cite{BP15}.

% \begin{thm}
%   {\upshape \cite{BP15}}
%   \label{the:char-conv}
%   Let $x \in \mathbb C^N$ be given by the convolution
%   \begin{equation}
%     \label{eq:def-conv}
%     x[n] = \left(x_1 * x_2\right) [n] 
%     := \sum_{k=0}^{N_1} x_1 [k] x_2[n-k],
%   \end{equation}
%   with the two signals $x_1 \in \mathbb C^{N_1}$ and
%   $x_2 \in \mathbb C^{N_2}$, where $N_1 \le N$ and $N_2 \le N$.  Then,
%   the signal $x'$ defined by 
%   \begin{equation*}
%     x'[n] 
%     := e^{j \alpha} (\overline x_1 [-\cdot]) * (x_2 [\cdot - n_0])
%     \qquad (\alpha \in [-\pi, \pi); n_0 \in \mathbb Z)
%   \end{equation*}
%   has the same Fourier intensity.  Moreover, for every signal $x'$ with
%   $\vert \hat x'(\omega) \vert = \vert \hat x(\omega) \vert$, there exists
%   two signals $x_1$, $x_2$ such that $x=x_1*x_2$ and
%   $x'=e^{j \alpha} (\overline x_1 [-\cdot]) * (x_2 [\cdot - n_0])$.
% \end{thm}

% {
% \begin{remark}
%   In the definition of the convolution \eqref{eq:def-conv}, only the
%   summands with $0 \le n-k < N_2$ are considered.  In other words,
%   we set $x_2 [n] = 0$ for $n < 0$ and $n \ge N_2$.
% \end{remark}
% }

\subsection{Ensuring uniqueness in classical phase retrieval}
\label{sec:ensuring-uniqueness}

To overcome the non-trivial ambiguities, and to ensure uniqueness in
the phase retrieval problem, one can rely on suitable a priori
conditions or further information about the true signal.  For
instance, if the sought signal represents an intensity or a
probability distribution, then it has to be real-valued and
non-negative.  Unfortunately, this natural contraint does not guarantee
uniqueness \cite{Bei16a}.  More appropriate priors
like minimum phase or sparsity ensure uniqueness for almost every or,
even, for every possible signal.  
Additional information about some entries of the true signal like the
magnitude or the phase also guarantee uniqueness in certain settings.
% Although the
% non-negativity usually reduces the number of feasible non-trivial
% solution, this additional information cannot be used to enforce
% uniqueness in general.  More precisely, the signals where the
% corresponding phase retrieval problem is unique as well as the signals
% which cannot be recovered uniquely up to trivial ambiguities form
% cones of infinite Lebesgue measure \cite{Bei16a}.

\subsubsection{Information about some entries of the true signal}
\label{sec:inform-about-some}

One approach to overcome the non-trivial ambiguities is to use
additional information about some entries of the otherwise unknown
signal $x$.  For instance, in wave front sensing and laser optics
\cite{SSD+06}, besides the Fourier intensity, the absolute values
$\vert x[0] \vert, \dots, \vert x[N-1] \vert$ of the sought signal $x$
are available.  Interestingly, already one absolute value
$\vert x[N-1-\ell] \vert$ within the support of the true signal $x$
almost always ensures uniqueness.

\begin{thm}
  \label{the:uni-abs-val}
  {\upshape\cite{BP16}} Let $\ell$ be an arbitrary integer between $0$
  and $N-1$.  Then almost every complex-valued signal
  $x\in \mathbb C^N$ with support length $N$ can be uniquely recovered
  from $y[k]$, $k=0,\dots, 2N-2$, in \eqref{eq:PR_classical} and
  $\vert x[N-1-\ell] \vert$ up to rotations if $\ell \ne (N-1)/2$.  In
  the case $\ell = (N-1)/2$, the reconstruction is almost surely
  unique up to rotations and conjugate reflections.
\end{thm}

% Although already one absolute value ensures almost always uniqueness,
% the knowledge of all absolute values
% $\vert x[0] \vert, \dots, \vert x[N-1] \vert$ of the true signal
% cannot surmount the `almost every' in Theorem~\ref{the:uni-abs-val}.
The uniqueness guarantee in Theorem~\ref{the:uni-abs-val} cannot be
improved by the knowledge of further or, even, all absolute values
$\vert x[0] \vert, \dots, \vert x[N-1] \vert$ of the true signal.
More precisely, one can explicitly construct signals that are not
uniquely defined by their Fourier magnitudes $y[k]$ and all temporal
magnitudes $\vert x[n] \vert$ for every possible signal length
\cite{BP16}.  In order to recover a signal from its Fourier magnitudes
and all temporal magnitudes numerically, several multi-level
Gauss-Newton methods have been proposed in \cite{LT08,LT09,SSD+06}.
Under certain  conditions, the convergence of these algorithms
to the true solution is guaranteed, and they allow signal
reconstruction from noise-free as well as from noisy data.

The main idea behind Theorem~\ref{the:uni-abs-val}
exploits $\vert x[N-1-\ell]\vert$ to show that the zero sets
$\{\beta_1, \dots, \beta_{N-1}\}$ of signals that cannot be recovered
uniquely (up to trivial ambiguities) form an algebraic variety of
lesser dimension.  This approach can be transferred to further kinds
of information about some entries of $x$.  For instance, the knowledge
of at least two phases of the true signal also guarantees uniqueness
almost surely.

\begin{thm}
  \label{the:uni-phase}
  {\upshape\cite{BP16}} Let $\ell_1$ and $\ell_2$ be different
  integers in $\{0, \dots, N-1\}$.  Then almost every complex-valued
  signal $x \in \mathbb C^N$ with support length $N$ can be uniquely
  recovered from $y[k]$, $k=0,\dots, 2N-2$, in
  \eqref{eq:PR_classical}, $\arg x[N-1-\ell_1]$, and
  $\arg x[N-1-\ell_2]$ whenever $\ell_1 + \ell_2 \ne N-1$.  In the
  case $\ell_1+\ell_2 = N-1$, the recovery is only unique up to
  conjugate reflection except for $\ell_1 = 0$ and $\ell_2 = N-1$,
  where the set of non-trivial ambiguities is not reduced at all.
\end{thm}

As a consequence of Theorems~\ref{the:uni-abs-val} and
\ref{the:uni-phase}, the classical phase retrieval problem is almost
always uniquely solvable if at least one entry of the true signal $x$
is known.  Unfortunately, there is no algorithm that knows how to
exploit the given entries to recover the complete signal in a stable
and efficient manner.

\begin{corollary}
  \label{cor:uni-entry}
  Let $\ell$ be an arbitrary integer between $0$ and $N-1$. Then
  almost every complex-valued signal $x \in \mathbb C^N$ with support
  length $N$ can be uniquely recovered from $y[k]$, $k=0,\dots, 2N-2$,
  in \eqref{eq:PR_classical} and $x [N-1-\ell]$ if $\ell \ne (N-1)/2$.
  In the case $\ell = (N-1)/2$, the reconstruction is almost surely
  unique up to conjugate reflection.
\end{corollary}

Corollary~\ref{cor:uni-entry} is a generalization of \cite{XYC87},
where the recovery of real-valued signals $x \in \mathbb R^N$ from
their Fourier magnitude $y[k]$ and one of their
end points $x[0]$ or $x[N-1]$ is studied.  In contrast to  Theorems~\ref{the:uni-abs-val} and
\ref{the:uni-phase}, the classical phase retrieval problem becomes
unique if enough entries of the true signal are known beforehand.

\begin{thm}
  \label{the:uni-L-end-pts}
  {\upshape\cite{BP15,nawab1983signal}} Each complex-valued signal
  $x\in \mathbb C^N$ with signal length $N$ is uniquely determined by
  $y[k]$, $k=0,\dots, 2N-2$, in \eqref{eq:PR_classical} and the
  $\lceil N/2 \rceil$ left end points
  $x[0], \dots, x[\lceil N/2 \rceil - 1 ]$.
\end{thm}

\subsubsection{Sparse signals}
\label{sec:sparse-signals}

In the last section, the true signal $x$ could be any arbitrary vector
in $\mathbb C^N$.  In the following, we consider the classical phase
retrieval problem under the assumption that the unknown signal is
sparse, namely, that only a small number of entries are non-zero.
Sparse signals have been studied thoroughly in the last two decades,
see for instance \cite{candes2006robust, donoho2006compressed,
  eldar2015sampling}.  Phase retrieval problems of sparse signals
arise in crystallography \cite{KH91, RCLV13} and astronomy \cite{BS79,
  RCLV13}, for example.  In many cases, the signal is sparse under an
unknown transform. In the context of phase retrieval, a recent paper
suggests a new technique to learn, directly from the phaseless data,
the sparsifying transformation and the sparse representation of the
signals simultaneously \cite{tillmann2016dolphin}.

The union of all $k$-sparse signals in $\mathbb C^N$, which have at
most $k$ non-zero entries, is here denoted by $\mathcal S_k^N$.  Since
$\mathcal S_k^N$ with $k<N$ is a $k$-dimensional submanifold of
$\mathbb C^N$ and hence itself a Lebesgue null set,
Theorem~\ref{the:uni-abs-val} and Corollary~\ref{cor:uni-entry} cannot
be employed to guarantee uniqueness of the sparse phase retrieval
problem.  Further, if the $k$ non-zero entries lie at equispaced
positions within the true signal $x$, i.e.,\ the support is of the form
$\{ n_0 + L m \colon m=0, \dots, k-1\}$ for some positive integers
$n_0$ and $L$, this specific phase retrieval problem is equivalent to
the recovery of a $k$-dimensional vector from its Fourier intensity
\cite{jaganathan2013sparse}.  Due to the non-trivial ambiguities,
which are characterized by Theorem~\ref{the:char-ambi}, the assumed
sparsity cannot always avoid non-trivial ambiguities.

In general, the knowledge that the true signal is sparse has a
beneficial effect on the uniquness of phase retrieval.  Under the
restriction that the unknown signal $x$ belongs to the class
$\mathcal T_k^N$ of all $k$-sparse signals in $\mathbb C^N$ without
equispaced support, which is again a $k$-dimensional submanifold, the
uniqueness is ensured for almost all signals.

\begin{thm}
  \label{the:uni-sparse}
  {\upshape\cite{jaganathan2013sparse}} Almost all signals
  $x \in \mathcal T_k^N$ can be uniquely recovered from their Fourier
  magnitudes $y[k]$, $k=0,\dots, 2N-2$, in \eqref{eq:PR_classical} up
  to rotations.
\end{thm}

Although Theorem~\ref{the:uni-sparse} gives a theoretical uniqueness
guarantee, it is generally a non-trivial task to decide whether a
sparse signal is uniquely defined by its Fourier intensity.  However,
if the true signal does not possess any collisions, uniqueness is
always given \cite{RCLV13}.  In this context, a sparse signal $x$ has
a \emph{collision} if there exists four indices $i_1$, $i_2$, $i_3$,
$i_4$ within the support of $x$ so that $i_1-i_2 = i_3 - i_4$.  A
sparse signal without collisions is called \emph{collision-free}.  For
instance, the signal
\begin{equation*}
  x = (0,0,1,0,-2,0,1,0,0,3,0,0)^T \in \mathbb R^{12}
\end{equation*}
is not collision-free since the index difference $6-4=2$ is equal to
$4-2=2$.

\begin{thm}
  \label{the:uni-no-col}
  {\upshape\cite{RCLV13}} Assume that the signal $x \in \mathcal S_k^N$ with
  $k<N$ has no collisions. 
  \begin{itemize}
  \item If $k \ne 6$, then $x$ can be uniquely recovered from $y[k]$,
    $k=0,\dots, 2N-2$, in \eqref{eq:PR_classical} up to trivial
    ambiguities;
  \item If $k=6$ and not all non-zero entries $x[n]$ have the same
    value, then $x$ can be uniquely recovered from $y[k]$,
    $k=0,\dots, 2N-2$, in \eqref{eq:PR_classical} up to trivial ambiguities;
  \item If $k=6$ and all non-zero entries $x[n]$ have the same value,
    then $x$ can be uniquely recovered from $y[k]$, $k=0,\dots, 2N-2$,
    in \eqref{eq:PR_classical} almost surely up to trivial
    ambiguities.
  \end{itemize}
\end{thm}

The uniqueness guarantees in Theorem~\ref{the:uni-no-col} remain valid
for $k$-sparse continuous-time signals, which are composed of $k$
pulses at arbitrary positions.  More precisely, the continuous-time
signal $f$ is here given by $f(t) = \sum_{i=0}^{k-1} c_i \, \delta(t-t_i)$,
where $\delta$ is the Dirac delta function, $c_i \in \mathbb C$ and
$t_i \in \mathbb R$.  In this setting, the uniqueness can be
guaranteed by $\mathcal O(k^2)$ samples of the Fourier magnitude
\cite{BP17}.

In Section~\ref{sec:sparse_algorithm}, we discuss different algorithms
to recover sparse signals $x \in \mathbb C^N$, that work well in practice.

\subsubsection{Minimum phase signals}
\label{sec:minim-phase-sign}

%Since the non-negativity of the true signal is not a suitable a priori
%condition to ensure uniqueness, we continue our literature review
%about uniqueness guarantees.  
Based on the observation that each
non-trivial solution of the classical phase retrieval problem is
uniquely characterized by the zero set
$\{\beta_1, \dots, \beta_{N-1}\}$ in Theorem~\ref{the:char-ambi}, one
of the simplest ideas to enforce uniqueness is to restrict these zeros
in an appropriate manner.  Under the assumption that the true signal
$x$ is a minimum phase signal, which means that all zeros $\beta_i$
chosen from the reflected zero pairs
$(\gamma_i, \overline \gamma_i^{\,-1})$ of the autocorrelation
polynomial $A(z)$ lie inside the unit circle, the corresponding phase
retrieval problem is uniquely solvable \cite{HLO80,haung2016}.

Although the minimum phase constraint  guarantees uniqueness,
the question arises how to ensure that an unknown signal is
minimum phase.  Fortunately, each complex-valued signal $x$ may be
augmented to a minimum phase signal.

\begin{thm}
  \label{the:min-phase}
  {\upshape\cite{haung2016}} For every $x \in \mathbb C^N$, the
  augmented signal
  \begin{equation*}
    x_{\mathrm{min}} = (\delta, x[0], \dots, x[N-1])^{T},
  \end{equation*}
  with $\vert \delta \vert \ge \| x \|_1$ is a minimum phase signal.
\end{thm}
Consequently, if the Fourier intensity of the augmented signal
$x_{\mathrm{min}}$ is available, then the true signal $x$ can always be
uniquely recovered up to trivial ambiguities.  Moreover,  the
minimum phase solution $x$ can be computed (up to rotations) from the
Fourier magnitude $y$ as in \eqref{eq:PR_classical} by a number of
efficient algorithms \cite{haung2016}.  Due to the trivial conjugate
reflection ambiguity, this approach  can be applied 
to maximum phase signals whose zeros lie outside the unit circle.

The minimum phase solution of a given phase retrieval problem may be
determined in a
stable manner using an approach by Kolmogorov 
\cite{haung2016}.  For simplicity, we restrict ourselves to the real
case $x \in \mathbb R^N$ with $x[N-1] > 0$.  The main idea is  to
determine the logarithm of the reversed polynomial
$\tilde X(z) = z^{N-1}\sum_{n=0}^{N-1} \overline x[n] \, z^{-n}$ from
the given data $y[k]$.  Under the assumption that all roots of $x$
strictly lie inside the unit circle, the analytic function
$\log \tilde X(z)$ may be written as
\begin{equation*}
  \log  \tilde X(z) = \sum_{n=0}^\infty \alpha_n \, z^{n},
  \qquad (\alpha_n \in \mathbb R)
\end{equation*}
where the unit circle $\vert z \vert = 1$ is contained in the region
of convergence.  Substituting $z = e^{-j \omega}$ with $\omega \in
\mathbb R$, we have
\begin{equation*}
  \Re \bigl[ \log \tilde X(e^{-j\omega}) \bigr] = \sum_{n=0}^\infty \alpha_n
  \cos \omega n
  \quad\text{and}\quad
  \Im \bigl[ \log \tilde X(e^{-j\omega}) \bigr] = - \sum_{n=0}^\infty \alpha_n
  \sin \omega n,
\end{equation*}
where $\Re[\cdot]$ and $\Im[\cdot]$ denote the real and imaginary parts, respectively.  
Since the real and imaginary part are a Hilbert transform pair,
$\Im \bigl[ \log \tilde X(e^{-j\omega}) \bigr]$ is completely defined
by $\Re \bigl[ \log \tilde X(e^{-j\omega}) \bigr]$. Because of the
identity
$\vert \tilde X (e^{-j\omega}) \vert^2= \vert A (e^{-j\omega}) \vert$,
the real part may be computed from the autocorrelation polynomial
$A(z)$ by
\begin{equation*}
   \Re \bigl[ \log \tilde X(e^{-j\omega}) \bigr] = \tfrac{1}{2} \log \vert A(e^{-j
   \omega}) \vert.
\end{equation*}
Finally, the autocorrelation polynomial $A(z)$ is completely
determined by the Fourier magnitudes $y[k]$, $k=0,\dots, 2N-2$,
leading to the recovery of the true minimum phase signal $x$.  % This
% reconstruction method can be generalized to complex-valued signal
% $x \in \mathbb C^N$, where the coefficients $\alpha_n$ may be complex.
Based on this idea, one can construct numerical algorithms that
guarantee stable signal recovery under the presence of noise
\cite{haung2016}.

\subsection{Phase retrieval with deterministic masks}
\label{sec:phase-retri-mask}

A further possibility to obtain additional information about the
underlying signal $x$ is to measure its Fourier magnitude with
respect to different masks as described in \eqref{eq:PR} and
\eqref{eq:PR_vector}.  Assuming that the masks are constructed
randomly, one can show that the corresponding phase retrieval problem
has a unique solution up to rotations almost surely or, at least, with
high probability. % \cite{BCM14,CLS13,Fan12,GKK16}.
Depending on the random model, the number of employed masks to recover
an one-dimensional signal $x \in \mathbb C^N$ varies form
$ O(\log N)$ over $ O((\log N)^2)$ to
$ O((\log N)^4)$, see \cite{BCM14}, \cite{GKK16}, and
\cite{CLS13} respectively.  Moreover, in the multidimensional case,
two independent masks are sufficient to guarantee uniqueness of almost
every signal up to rotations \cite{Fan12}.  As the following results
show, in the deterministic setup, already a very small number of
specifically constructed masks ensure uniqueness for
most signals.

% The random approach leads to two main issues in applications.
% Firstly, how can a mask be constructed with respect to a certain
% random distribution, and secondly, how can the uniqueness for a set of
% specific masks be verified.  In the case where only two deterministic
% masks are involved, the second question can be answered as follows.

\begin{thm}
  \label{the:uni-arb-masks}
  {\upshape\cite{jaganathan2015phase_mask}} Almost all complex-valued
  signals $x \in \mathbb C^N$ can be uniquely recovered from $y[m,k]$,
  $m=1,2$ and $k=0,\dots, 2N-2$, as in \eqref{eq:PR} up to rotations
  if the two masks $d_1$, $d_2 \in \mathbb C^N$ satisfy
  \begin{itemize}
  \item $d_1 [n] \ne 0$ or $d_2 [n] \ne 0$ for each $0 \le n
    \le N-1$,
  \item $d_1 [n] d_2 [n] \ne 0$ for some $0 \le n \le N-1$.
  \end{itemize}
\end{thm}

For some masks $d_1$ and $d_2$, one can overcome the `almost
all' in Theorem~\ref{the:uni-arb-masks} and obtain uniqueness of
the corresponding phase retrieval problem.

\begin{thm}
  \label{the:uni-two-masks}
  {\upshape\cite{jaganathan2015phase_mask}}
  If the diagonal matrices $D_1$, $D_2$ correspond to the two masks
  \begin{equation} \label{eq:fixed_masks}
    d_1[n] = 1 \quad (0 \le n \le N-1)
    \qquad\text{and}\qquad
    d_2[n] =
    \begin{cases}
      0 & n=0 \\
      1 & 1 \le n ,\le N-1,
    \end{cases}
  \end{equation}
  then every complex-valued signal $x \in \mathbb C^N$ with
  $x[0] \ne 0$ can be uniquely recovered from $y[m,k]$, 
  $m=1,2$  and $k=0,\dots, 2N-2$, up to rotations.
\end{thm}

% \begin{thm}
%   \label{the:uni-five-masks}
%   {\upshape\cite{jaganathan2015phase_mask}} Assume that the signal
%   length $N$ of the true signal is a multiple of four.  If the diagonal
%   matrices $D_1, \dots, D_5$ correspond to the five masks
%   \begin{align*}
%     d_1[n] &= \begin{cases}
%       1 & 0 \le n \le N/2 -1 \\
%       0 & N/2 \le n \le N-1, \\
%     \end{cases} & 
%     d_2[n] &= \begin{cases}
%       0 & n=0 \\
%       1 & 1 \le n \le N/2-1 \\
%       0 & N/2 \le n \le N-1 \\
%     \end{cases}
%     \\
%     d_3 [n] &= \begin{cases}
%       0 & 0 \le n \le N/2 \\
%       1 & N/2 +1 \le n \le N-1\\
%     \end{cases} &
%     d_4 [n] &= \begin{cases}
%       0 & \le n \le N/2 -1 \\
%       1 & N/2 \le n \le N-1 \\
%     \end{cases}
%     \\
%     d_5 [n] &= \begin{cases}
%       0 & 0 \le n \le N/4 -1 \\
%       1 & N/4 \le n \le 3N/4-1 \\
%       0 & 3N/ 4 \le n \le N-1,
%     \end{cases}
%   \end{align*}
%   then every complex-valued signal $x \in \mathbb C^N$ with
%   $x[0] \ne 0$, $x[N/2-1] \ne 0$, and $x[N/2] \ne 0$ can be uniquely
%   recovered from $\vert \widehat{D_\ell x} (\omega) \vert$ with
%   $1\le \ell \le 5$ up to rotations. 
% \end{thm}

A different approach to exploit deterministic masks in order to
overcome the ambiguity in phase retrieval is discussed in
\cite{JH16} and can be proven by using the characterization in
Theorem~\ref{the:char-ambi}.  More explicitly, here the two masks
\begin{equation}
  \label{eq:mask-auto-cross}
  d_1[n] =
  \begin{cases}
    1, & 0 \le n \le L-1, \\
    0, & L \le n \le N-1, \\
  \end{cases}
  \qquad\text{and}\qquad
  d_2[n] =
  \begin{cases}
    0, & 0 \le n \le L-1,\\
    1, & L \le n \le N-1, \\
  \end{cases}
\end{equation}
for some $L$ between $1$ and $N-2$ are used.  Pictorially, the mask
$d_1$ blocks the right-hand side of the underlying signal $x$ and
$d_2$ the left-hand side.

For the signals $x$, $D_1 x$, and $D_2 x$, where $D_i$ is the diagonal
matrix with respect to the mask $d_i$, we  define the polynomials
$X$, $X_1$, and $X_2$ by
\begin{equation*}
  X(z) = \sum_{n=0}^{N-1} x[n] \, z^n, 
  \quad
  X_1 (z) = \sum_{n=0}^{L-1} x[n] \, z^n
  \quad\text{and}\quad
   X_2 (z) = \sum_{n=1}^{N-L-1} x[n+L] \, z^n.
 \end{equation*}
 Different from the autocorrelation functions of $D_1 x$ and
 $D_2 x$, which are simply given by
 $A_1(z) = X_1(z) \tilde X_1(z)$ and
 $A_2(z) = X_2(z) \tilde X_2(z)$, the autocorrelation function $A(z)$
 of the true signal $x$ can be written as
\begin{align*}
  A (z)
  &= \bigl(X_1(z) + z^L X_2(z)\bigr)
    \bigl( z^{N-L-1} \tilde X_1(z) + \tilde
    X_2(z)\bigr)
  \\
  &=z^{N-L-1} A_1 (z) + X_1(z) \tilde
    X_2(z)
    + z^{N-1} \tilde X_1 (z) X_2(z)
    + z^L A_2(z),
\end{align*}
since $X(z) = X_1(z) + z^L X_2(z)$.  Due to the fact that
$X_1(z) \tilde X_2(z)$ and $ z^{N-1} \tilde X_1 (z) X_2(z)$ have no
common monomials with the same degree, one can determine the
polynomials
\begin{equation}
  \label{eq:auto-cross-poly}
  X_1(z) \tilde X_1(z),
  \quad
  X_1(z) \tilde X_2(z),
  \quad
  \tilde X_1(z) X_2(z),
  \quad\text{and}\quad
  X_2(z) \tilde X_2(z),
\end{equation}
from the autocorrelation functions $A (z)$,
$A_1 (z)$ and $A_2 (z)$.

As mentioned before, the reversed polynomials $\tilde X_i(z)$
correspond to the reflected zero set of $X_i (z)$ with respect to the
unit circle.  Hence, assuming that the zeros of $D_1 x$ and $D_2 x$
are pairwise different, one can determine both zero sets by comparing
the roots of the four polynomials \eqref{eq:auto-cross-poly}, which
yields the following result.

\begin{thm}
  \label{the:uni-auto-cross-cor}
  {\upshape\cite{JH16}} Let $x \in \mathbb C^N$, and assume that the
  zeros $\xi_i$ and $\eta_\ell$ of 
  \begin{align*}
    X_1 (z) 
    = x[L-1] \prod_{i=1}^{L-1} (z - \xi_i),
    \quad\text{and}\quad
    X_2 (z)
    = x[N-1] \prod_{\ell=1}^{N-L-1} (z - \eta_\ell),
  \end{align*}
  are pairwise different.  Then the signal $x$ can be uniquely
  recovered up to rotations from the Fourier magnitudes $y[m,k]$,
  $m=0,1,2$ and $k=0,\dots, 2N-2$, with the masks $d_0 \equiv 1$ and
  $d_1$, $d_2$ in \eqref{eq:mask-auto-cross}.
\end{thm}

The phase retrieval problem in Theorem~\ref{the:uni-auto-cross-cor} is
equivalent to the recovery of $x_1=D_1 x$ and $x_2=D_2 x$ with support
$\{0,\dots, L-1\}$ and $\{L, \dots, N-1\}$ from the Fourier magnitudes
of $x_1$, $x_2$, and $x_1 + x_2$.
% $\vert \widehat{D_1 x} (\omega) \vert$,
% $\vert \widehat{D_2 x} (\omega) \vert$, and
% $\vert \widehat{D_1 x} (\omega) + \widehat{D_2 x} (\omega) \vert$.
More generally, the recovery of two arbitrary signals $x_1$,
$x_2 \in \mathbb C^N$ from their Fourier magnitudes and the Fourier
magnitude of the interference $x_1+x_2$ has been studied in
\cite{BP15,KH93}. Theorem~\ref{the:uni-auto-cross-cor} is a
specific instance of the uniqueness guarantee given in \cite{BP15}.
Furthermore, these problems are closely related to the vectorial phase
retrieval problem introduced in \cite{leshem2017discrete,RDN13,raz2011vectorial}, where the
Fourier magnitudes of a second interference $x_1 + j x_2$ are employed.

% To solve the vectorial phase retrieval problem numerically, The algorithm is based on the minimization of a
% quadratic functional and an estimation to determine the unknown
% supports \cite{RDN13}.

% Where the numerical recovery in the general case is difficulty, the
% phase retrieval problem in Theorem~\ref{the:uni-auto-cross-cor} can be
% solved by a semidefinite programming-based algorithm that is stable
% under noise \cite{JH16}.  

A further example for phase retrieval with deterministic masks is considered in
\cite{candes2015phase}, where the three masks are defined by
\begin{equation}
  \label{eq:mod-mask}
  d_0 [n] = 1,
  \qquad
  d_1[n] = 1 + e^{2 \pi j s n/N},
  \quad\text{and}\quad
  d_2[n] = 1 + e^{2 \pi j (sn / N - 1/4)},
\end{equation}
for a non-negative integer $s$.  The masks $d_1$ and $d_2$ here
interfere the unknown signal $x$ with a modulated version of the
unknown signal itself, which yields the Fourier magnitudes
$\vert \hat x [k] + \hat x [k-s] \vert^2$ and
$\vert \hat x [k] - j\, \hat x [k-s] \vert^2$.  Together with the
Fourier magnitudes $\vert \hat x [k] \vert^2$, for almost every
signal, the relative phases $\phi [k-s] - \phi [k]$ of the Fourier
transform $\hat x [k] = \vert x [k] \vert \, e^{j \phi [k]}$ can be
determined.  If $s$ is relatively prime with $N$, then the Fourier transform
$\hat x$ and thus the true signal $x$ are recovered up to
rotations.

\begin{thm}
  \label{the:mod-mask}
  {\upshape\cite{candes2015phase}}
  Let $x \in \mathbb C^N$ be a signal with non-vanishing  DFT.  Then
  $x$ is uniquely recovered from $y[m,k]$ with $K=\tilde N = N$
  and the masks in \eqref{eq:mod-mask} up to rotations if and
  only if the non-negative integer $s$ is relatively prime with $N$.
\end{thm}

The masks in \eqref{eq:mod-mask} as well as the uniqueness guarantee
in Theorem~\ref{the:mod-mask} can be generalized to multidimensional
phase retrieval \cite{candes2015phase}.  If $\tilde N$ is replaced by
$2N-1$, every signal $x \in \mathbb C^N$ is uniquely recovered up
to rotation from its Fourier magnitudes $y[m,k]$ in \eqref{eq:PR} with
masks $d_0[n] = 1$ and
$d_i[n] = 1 + e^{j\alpha_i} \, e^{2 \pi j s n/N}$, $i=1,2$, where
$\alpha_i \in [-\pi, \pi)$, and where $s$ can be nearly every real
number \cite{Bei16}.  Several further examples of deterministic masks
which allow a unique recovery are detailed in \cite{Bei16,
  candes2015phase,jaganathan2015phase_mask, JH16} and references
therein.  In Section~\ref{sec:sdp}, we consider semidefinite
relaxation algorithms which stably recover the unknown signal from its
masked Fourier magnitudes \eqref{eq:PR} under noise.

\subsection{Phase retrieval from STFT measurements}
\label{sec:phase-retri-stft}

We next consider uniqueness guarantees for the
recovery of an unknown signal from the magnitude of its STFT as defined in \eqref{eq:PR_STFT}.  This problem
can be interpreted as a sequence of classical phase retrieval
problems, where some entries of the underlying signals have to coincide.  Obviously, the STFT phase
retrieval problem cannot be solved uniquely if the parameter $L$ is
greater than or equal to the window length $W$, since the classical
problems are then independent from each other.

Under the assumption
that the known window $d$ does not vanish, i.e.,\ $d[n] \ne 0$ for
$n=0, \dots, W-1$, some of the first uniqueness guarantees were
established in \cite{nawab1983signal}.

% \begin{thm}
%   {\upshape\cite{nawab1983signal}}
%   $x \in \mathbb R^N$ uniquely defined: $L=1$ 

%   window: Length $N_w>1$, no zeros within the support

%   signal: at most $N_w-2$ consecutive zero samples between any two
%   non-zero samples, sign of first non-zero sample known. 
% \end{thm}

\begin{thm}
  \label{the:stft-nawab}
  {\upshape\cite{nawab1983signal}} Let $d$ be a non-vanishing window
  of length $W > 2$, and let $L$ be an integer in
  $\{1, \dots, \lfloor W/2 \rfloor \}$.  If the signal
  $x \in \mathbb C^N$ with support length $N$ has at most $W - 2L$
  consecutive zeros between any two non-zero entries, and if the first
  $L$ entries of $x$ are known, then $x$ can be uniquely
  recovered from $y[m,k]$ with $K = 2W-1$ in \eqref{eq:PR_STFT}.
\end{thm}

The main idea behind Theorem~\ref{the:stft-nawab} is that the
corresponding classical phase retrieval problems are solved
sequentially.  For instance, the case $m=1$ is equivalent to recovering a
signal in $\mathbb C^{L+1}$ from its Fourier intensity and the first
$L$ entries.  The uniqueness of this phase retrieval problem is
guaranteed by Theorem~\ref{the:uni-L-end-pts}.  Since the true signal
$x$ has at most $W-2L$ consecutive zeros, the remaining subproblems
can also be reduced to the setting considered in
Theorem~\ref{the:uni-L-end-pts}.

Knowledge of the first $L$ entries of $x$ in
Theorem~\ref{the:stft-nawab} is a strong restriction in practice.
Under the a priori constraint that the unknown signal is non-vanishing
everywhere, the first $L$ entries are not needed to ensure uniqueness.

\begin{thm}
  \label{the:stft-almost-uni}
  {\upshape\cite{jaganathan2016stft}} Let $d$ be a non-vanishing
  window of length $W$ satisfying $L < W \le N/2$.  Then almost all
  non-vanishing signals can be uniquely recovered up to rotations from
  their STFT magnitudes $y[m,k]$ in \eqref{eq:PR_STFT} with $2W \le K
  \le N$ and $\tilde N = N$.
\end{thm}

For some classes of STFT windows, the uniqueness is
guaranteed for all non-vanishing signals
\cite{bendory2016non,eldar2015sparse}.  Both references use a slightly
different definition of the STFT, where the
STFT window in \eqref{eq:PR_STFT} is periodically extended over the
support $\{0, \dots, N-1\}$, i.e.,\ the indices of the window $d$ are
considered as modulo the signal length $N$.

\begin{thm}
  \label{the:per-stft-uni:1}
  {\upshape\cite{eldar2015sparse}} Let $d$ be a periodic window with
  support length $W \ge 2$ and $2W-1 \le N$, and assume that the
  length-$N$ DFT of $\vert d[n] \vert^2$ is non-vanishing.  If $N$ and
  $W-1$ are co-prime, then every non-vanishing signal
  $x \in \mathbb C^N$ can be uniquely recovered from its STFT
  magnitudes $y[m,k]$ in \eqref{eq:PR_STFT} with $L=1$ and
  $K=\tilde N = N$ up to rotations.
\end{thm}

\begin{thm}
  \label{the:per-stft-uni:2}
  {\upshape\cite{bendory2016non}} Let $d$ be a periodic window of
  length $W$, and assume that the length-$N$ DFT of
  $\vert d[n] \vert^2$ and $d[n] d[n-1]$ are non-vanishing.  Then
  every non-vanishing signal $x \in \mathbb C^N$ can be uniquely
  recovered from its STFT magnitudes $y[m,k]$ in \eqref{eq:PR_STFT}
  with $L=1$ and $K=\tilde N = N$ up to rotations.
\end{thm}

If we abandon the constraint that the underlying signal is
non-vanishing, then the behaviour of the STFT phase retrieval problem
changes dramatically, and the recovery of the unknown signal becomes
much more challenging.  For example, if the unknown signal $x$
possesses more than $W-1$ consecutive zero entries, then the signal
can be divided in two parts, whose STFTs are completely independent.
An explicit non-trivial ambiguity for this specific setting is
constructed in \cite{eldar2015sparse}.  Depending on the window
length, there are thus some natural limitations on how far uniqueness can
be ensured for sparse signals.
% Again, a first uniqueness guarantee is given by
% Theorem~\ref{the:stft-nawab}, which is not restricted to non-vanishing
% signals.  Depending on the window length $W$ and the parameter $L$,
% the unknown signal can here have up to $W-2L$ consecutive zeros.
% Exploiting that overlapping and
% non-overlapping parts of sufficient many adjacent section contain at
% least one non-zero entry of the underlying signal, Theorem~\ref{the:stft-almost-uni} can be generalized for sparse
% signals. 

\begin{thm}
  \label{the:stft-uni-sparse}
  {\upshape\cite{jaganathan2016stft}} 
  Let $d$ be a non-vanishing window of length $W$ satisfying $L<W\le
  N/2$.  Then almost all sparse signals with less than $\min\{W-L,
  L\}$ consecutive zeros can be uniquely recovered up to rotations
  from their STFT magnitudes $y[m,k]$ in \eqref{eq:PR_STFT} with $2W
  \le K \le N$ and $\tilde N = N$.
\end{thm}

In \cite{bojarovska2016phase}, the STFT is interpreted as measurements
with respect to a Gabor frame.  Under certain conditions on the
generator of the frame, every signal $x \in \mathbb C^N$ is uniquely
recovered up to rotations. Further, the true signal $x$ is given as a
closed form solution.  For the STFT model in
\eqref{eq:PR_STFT}, this implies the following uniqueness guarantee.

\begin{thm}
  \label{the:gabor-frame}
  {\upshape\cite{bojarovska2016phase}} Let $d$ be a periodic window of
  length $W$, and assume that the length-$N$ DFT of
  $d[n] d[n-m]$ is non-vanishing for $m=0,\dots,N-1$.  Then
  every signal $x \in \mathbb C$ can be uniquely
  recovered from its STFT magnitudes $y[m,k]$ in \eqref{eq:PR_STFT}
  with $L=1$ and $K=\tilde N = N$ up to rotations.
\end{thm}

The main difference between Theorem~\ref{the:gabor-frame} and the
uniqueness results before is that the unknown signal
$x \in \mathbb C^N$ can have arbitrarily many consecutive zeros.  On the
other hand, the STFT window must have a length of at least $N/2$ in
order to ensure that $d[n] d[n-m]$ is not the zero vector.  Thus, the
thm is only relevant for long windows.  A similar result was
derived in \cite{bendory2016non}, followed by a stable recovery
algorithm; see Section~\ref{sec:stft_alg}.

% \begin{thm}
%   \label{the:admiss-window}
%   {\upshape\cite{bendory2016non}} Let $d$ be a periodic admissible
%   window of length $W \ge \lceil (N+1)/2 \rceil$, Then every signal
%   $x \in \mathbb C$ can be uniquely recovered from its STFT magnitudes
%   $y[m,k]$ in \eqref{eq:PR_STFT} with $L=1$ and $K=\tilde N = N$ up to
%   rotations.
% \end{thm}

\subsection{FROG methods }  \label{sec:FROG}

An important optical application for phase retrieval is ultra-short laser pulse characterization \cite{trebino2012frequency, trebino1997measuring}. One way to overcome the non-uniqueness of Fourier phase retrieval in this application is by employing  
a measurement technique called  X-FROG (see also Section \ref{sec:problem_formulation}). In X-FROG, a reference window is used to gate the sought signal, resulting in the STFT phase retrieval model (\ref{eq:PR_STFT}).
However, in practice it is quite hard to generate and measure such a reference window. Therefore, in order to generate redundancy
in ultra-short laser pulse measurements it is common to  correlate the signal with itself. This method is called Frequency-Resolved Optical Gating
(FROG). 

\begin{figure*}[t]
  \centering
  {\includegraphics[scale=0.6]{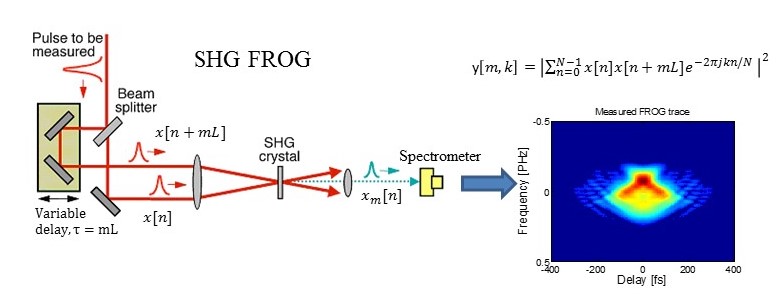}}
  \caption{\label{fig:FROG} Illustration of the SHG FROG technique (courtesy of \cite{bendory2017on}).}
\end{figure*}

FROG is probably the most
commonly used approach for full characterization of ultra-short
optical pulses due to its simplicity and good experimental performance.
  A FROG apparatus
produces a 2D intensity diagram of an input pulse by interacting the
pulse with delayed versions of itself in a nonlinear-optical medium,
usually using a second harmonic generation (SHG) crystal
\cite{delong1994frequency}. This 2D signal is called a FROG trace and
is a quartic function of the unknown signal.  An illustration of the
FROG setup is presented in Fig.~\ref{fig:FROG}. Here we
focus on SHG FROG but other types of nonlinearities exist for FROG
measurements. A generalization of FROG, in which two different unknown
pulses gate each other in a nonlinear medium, is called blind
FROG. This method can be used to characterize simultaneously two
signals \cite{wong2012simultaneously}. In this case, the measured data
is referred to as a blind FROG trace and is quadratic in both signals.
We refer to the problems of recovering a signal from its blind FROG
trace and FROG trace as \emph{bivariate phase retrieval} and
\emph{quartic phase retrieval}, respectively.

In bivariate phase retrieval we acquire, for each delay step $m$, the
power spectrum of
\begin{equation*} 
  {x}_{m}[n]={x}_1\left[n\right]{{x}_2\left[n+mL\right]},
\end{equation*}
where $L$, as in the STFT phase retrieval setup, determines the
overlap factor between adjacent sections.  The acquired data is given
by
\begin{equation} \label{eq:FROG}
  \begin{split}
    {y}\left[m,k\right]&=\left\vert \sum_{n=0}^{N-1}{x}_{m}\left[n\right]e^{-2\pi jkn/N}\right\vert^2  \\
    &= \left\vert \sum_{n=0}^{N-1} {x}_1\left[n\right]{{x}_2\left[n+mL\right]}e^{-2\pi jkn/N}\right\vert^2.
  \end{split}
\end{equation}
Quartic phase retrieval is the special case in which
${x}_1={x}_2$.

The trivial ambiguities of bivariate phase retrieval are described in
the following proposition.

\begin{proposition}
  \label{prop:ambiguities}
  {\upshape\cite{bendory2017on}} Let ${x}_1,{x}_2\in\mathbb{C}^N$ and
  let ${x}_m[n]:={x}_1[n]{x}_2[n+mL]$ for some fixed $L$. Then, the
  following signals have the same phaseless bivariate measurements
  $y[m,k]$ as the pair ${x}_1,{x}_2$:
  \begin{enumerate}
  \item multiplication by {global phases} $ {x}_1e^{j\psi_1},{x}_2e^{j\psi_2}$
    for some $\psi_1,\psi_2\in\mathbb{R}$,
  \item the {shifted} signal ${\tilde{x}}_m[n]={{x}}_m[n-n_0]$
    for some $n_{0}\in\mathbb{Z}$,
  \item the {conjugated and reflected} signal ${\grave{x}}_m[n]=\overline{{{{x}}}_m[-n]}$,
  \item modulation, ${x}_1[n]e^{-2\pi jk_0n/N}$, ${x}_2[n]e^{2\pi jk_0n/N}$ for some $k_0\in\mathbb{Z}$.
  \end{enumerate}
\end{proposition}

The fundamental question of uniqueness for FROG methods has been
analyzed first in \cite{seifert2004nontrivial} for the continuous
setup. The analysis of the discrete setup appears in~\cite{bendory2017on}.

\begin{thm}
  \label{thm:main_frog}
  {\upshape\cite{bendory2017on}} Let $L=1$, and let $\hat{{x}}_1$ and
  $\hat{{x}}_2$ be the Fourier transforms of ${x}_1$ and ${x}_2$,
  respectively. Assume that $\hat{{x}}_1$ has at least
  $\left\lceil (N-1)/2\right\rceil$ consecutive zeros
  (e.g.,\ band-limited signal).  Then, almost all signals are determined
  uniquely, up to trivial ambiguities, from the measurements
  ${y}[m,k]$ in \eqref{eq:FROG} and the knowledge of
  $\vert \hat{{x}}_1\vert $ and $\vert \hat{{x}}_2\vert $. By trivial
  ambiguities we mean that ${x}_1$ and ${x}_2$ are determined up to
  global phase, time shift and conjugate reflection.
\end{thm}

 Several heuristic techniques have been proposed to
estimate an underlying signal from its FROG trace.
These algorithms are based on a variety of methods, such as
alternating projections, gradient descent and iterative PCA
\cite{kane2008principal, sidorenko2016ptychographic, trebino1993using}.

%
%To provide a complete picture, we describe other FROG nonlinearities. Two
%examples are third-harmonic generation FROG and polarization gating
%FROG in which  measured signal is modeled as the
%Fourier magnitudes of ${x}_m[n] ={x}^2[n]{x}[n-mL]$ and
%${x}_m[n] ={x}[n]\vert {x}[n-mL]\vert$, respectively
%\cite{trebino1997measuring, tsang1996frequency}.  Another important
%application is the so called Frequency-Resolved Optical Gating for
%Complete Reconstruction of Attosecond Bursts (FROG CRAB), which is
%based on the photoionization of atoms by the attosecond field. In this setup, the signal is
%modeled as the Fourier magnitudes of
%${x}_m[n] ={x}_1[n]e^{j{x}_2[n-mL]}$ \cite{mairesse2005frequency}.
%

%%% Local Variables:
%%% mode: latex
%%% TeX-master: "draft4"
%%% End:

\subsection{Multidimensional phase retrieval}
\label{sec:multidimens-phase}

In a wide range of real-world
applications like crystallography or electron microscopy, the natural
objects of interest correspond to two or three-dimensional signals.
% Hence, the question arises whether the results about one-dimensional
% phase retrieval can be extended to the multidimensional case.
More generally, the $r$-dimensional
phase retrieval problem consists of the recovery of an unknown
$r$-dimensional signal $x \in \mathbb C^{N_1 \times \cdots \times
  N_r}$ from its Fourier magnitudes
\begin{equation}
  \label{eq:multi-four-mag}
  \begin{split}
    &y[k] = \left\vert \sum_{n \in \mathbb Z_N} x[n] e^{-(2\pi)^r j \,
        k \cdot n / \tilde N_1\dots \tilde N_r} \right\vert^2 \!,
    \\[10pt]
    &k \in
    \{0, \dots,  K_1-1 \} \times \cdots \times \{0, \dots,
     K_r-1 \},
  \end{split}
\end{equation}
with $n=(n_1, \dots, n_r)^T$ and
$\mathbb Z_N = \{0, \dots, N_1-1 \} \times \cdots \times \{0, \dots,
N_r-1 \}$.
Unless otherwise mentioned, we assume $\tilde N_i = K_i = 2N_i - 1$.

Clearly, rotations, transitions, or conjugate reflections of the true
signal lead to trivial ambiguities.  Besides these similarities, the
ambiguities of the multidimensional phase retrieval problem are very
different from those of its one-dimensional counterpart.  More
precisely, non-trivial ambiguities occur only in very rare cases, and
almost every signal is uniquely defined by its Fourier magnitude up to
trivial ambiguities.

% If we look back to the one-dimensional setting, the main reason for
% the occurring non-trivial ambiguities is that we cannot decide whether
% the factor $(e^{-j \omega} - \gamma_i)$ or
% \raisebox{0pt}[0pt][0pt]{$(e^{-j \omega} - \overline
%   \gamma_i^{\,-1})$}
% of the autocorrelation function $\hat a$ in \eqref{eq:auto-fun-1d}
% belongs to the true signal $x$.  An analogous problem arises for the
% multidimensional case \cite{Hay82}.  Here, the Fourier transform of a
% $r$-dimensional signal
% $x \in \mathbb C^{N_1 \times \cdots \times N_r}$ is the restriction of
% a multivariate algebraic polynomial $X(z)$ with $z = z_1, \dots, z_r$
% to the unit circle, i.e.
% \begin{equation*}
%   \hat x (\omega) = X(e^{-j \omega}),
% \end{equation*}
% with $e^{-j \omega} = (e^{-j \omega_1}, \dots, e^{-j \omega_r})^T$.

Similarly to Section~\ref{sec:trivial-non-trivial}, the non-trivial
ambiguities can be characterized by exploiting the autocorrelation.
Here the related polynomial
\begin{equation*}
  X(z) = \sum_{n \in \mathbb Z_n} x[n] \, z^n
   = \prod_{i=1}^I X_i (z),
\end{equation*}
with $z^n = z^{n_1} \cdots z^{n_r}$ is uniquely factorized (up to
multiplicative constants) into irreducible factors $X_i(z)$, which
means that the $X_i$ cannot be represented as a product of
multivariate polynomials of lesser degree.  The main difference with the
one-dimensional setup is that most multivariate polynomials consists
of only one irreducible factor $X_i$.
% Since the multivariate polynomials form a factorial ring, the
% polynomial $X$ can be uniquely written (up to multiplicative constants) as
% \begin{equation*}
%   X(z) = \prod_{i=1}^I X_i (z),
% \end{equation*}
% with the non-trivial irreducible factors $X_i$. 
Denoting the multivariate reversed
polynomial  by
\begin{equation*}
  \tilde X_i (z) = z^{M} \, \overline X_i(z^{\,-1}),
\end{equation*}
with $z^M = z^{M_1} \cdots z^{M_r}$, where $M_\ell$ is the degree of
$X_i$ with respect to the variable $z_\ell$, the non-trivial ambiguities in the
multidimensional setting are characterized as follows.
% , and where $\overline X_i$
% is the polynomial with conjugate coefficients,
% we can finally write
% the squared Fourier intensity as
% \begin{align*}
%   \hat a (\omega) 
%   = \vert \hat x (\omega) \vert^2
%   &= \prod_{i=1}^I \vert X_i (e^{-j \omega}) \vert \, \vert \overline X_i (e^{j
%   \omega}) \vert
%   \\
%   &= \prod_{i=1}^I \vert X_i (e^{-j\omega}) \vert \, \vert \tilde
%     X_i (e^{-j \omega}) \vert.
% \end{align*}
% Based on this factorization, the non-trivial ambiguities in the
% multidimensional setting can be characterized as follows.

\begin{thm}
  \label{the:char-multi}
  {\upshape\cite{Hay82}} Let
  $x \in \mathbb C^{N_1 \times \cdots \times N_r}$ be the
  complex-valued signal related to the polynomial
  $X(z) = \prod_{i=1}^I X_i (z)$,
  where $X_i(z)$ are non-trivial irreducible polynomials.  Then the
  polynomial $X'(z)=\sum_{n \in \mathbb Z_N} x'[n] \, z^n$ of each signal
  $x' \in \mathbb C^{N_1 \times \cdots \times N_r}$ with
  Fourier magnitudes $y'[k] = y[k]$ in \eqref{eq:multi-four-mag} can
  be written as
  \begin{equation*}
    X' (z) = \prod_{i \in J} X_i (z) \cdot \prod _{i \notin J}
    \tilde X_i (z),
  \end{equation*}
  for some index set $J \subset \{1, \dots, I \}$.
\end{thm}

Thus, the phase retrieval
problem is uniquely solvable up to trivial ambiguities if the algebraic polynomial $X(z)$ of the true signal $x$ is
irreducible, or if all but one factor $X_i(z)$ are invariant under
reversion \upshape\cite{Hay82}.  In contrast to the one-dimensional
case, where the polynomial $X(z)$ may always be factorized into linear
factors with respect to the zeros $\beta_i$, cf.\
Theorem~\ref{the:char-ambi}, most multivariate polynomials cannot be
factorized as mentioned above.

\begin{thm}
  \label{the:irred-poly}
  {\upshape\cite{HM82}} The subset of the $r$-variate polynomials
  $X(z_1, \dots, z_r)$ with $r>1$ of degree $M_\ell>1$ in $z_\ell$
  which are reducible over the complex numbers corresponds to a set of
  measure zero.
\end{thm}

Consequently, the multidimensional phase retrieval problem has a
completely different behaviour than its one-dimensional counterpart.

\begin{corollary}
  \label{cor:uni-multi}
  Almost every $r$-dimensional signal with $r>1$ is uniquely defined
  by its Fourier magnitudes $y[k]$ in \eqref{eq:multi-four-mag} up to
  trivial ambiguities.
\end{corollary}

Investigating the close connection between the one-dimensional and
two-di\-men\-sion\-al problem formulations, the different uniqueness
properties have been studied in \cite{kogan20162d}.  Particularly, one
can show that the two-dimensional phase retrieval problem corresponds
to a one-dimensional problem with additional constraints, which
almost always guarantee uniqueness.
Despite these uniqueness guarantees, there
are no systematic methods to estimate an $r$-dimensional signal from
its Fourier magnitude \cite{bauschke2002phase,kogan20162d}. The most
popular techniques are based on alternating projection algorithms as
discussed in Section \ref{sec:alternating_projection}.

%%% Local Variables:
%%% mode: latex
%%% TeX-master: "draft4"
%%% End:

\section{ Phase retrieval algorithms} \label{sec:algorithms}

The previous section presented conditions under which there exists a
unique mapping between a signal and its Fourier magnitude (up to
trivial ambiguities). Yet, the existence of a unique mapping does not
imply that we can actually estimate the signal in a stable fashion. The goal of this
section is to present different algorithmic approaches for the inverse
problem of recovering a signal from its phaseless Fourier
measurements.  In the absence of noise, this task can be formulated as
a feasibility problem over a non-convex set
\begin{equation} \label{eq:pr_feas}
  \begin{split} 
    \mbox{find}_{z\in\mathbb{C}^N}\quad\mbox{subject to}\quad &y[m,k]=\left\vert f_k^*D_mz\right\vert^2 ,\\ & k=0,\dots,K-1, \quad m=0,\dots,M-1.
  \end{split}
\end{equation}
Recall that \eqref{eq:pr_feas} covers the classical and STFT phase retrieval problems as special cases.

\begin{figure} 
  \centering
  \includegraphics[scale=0.25]{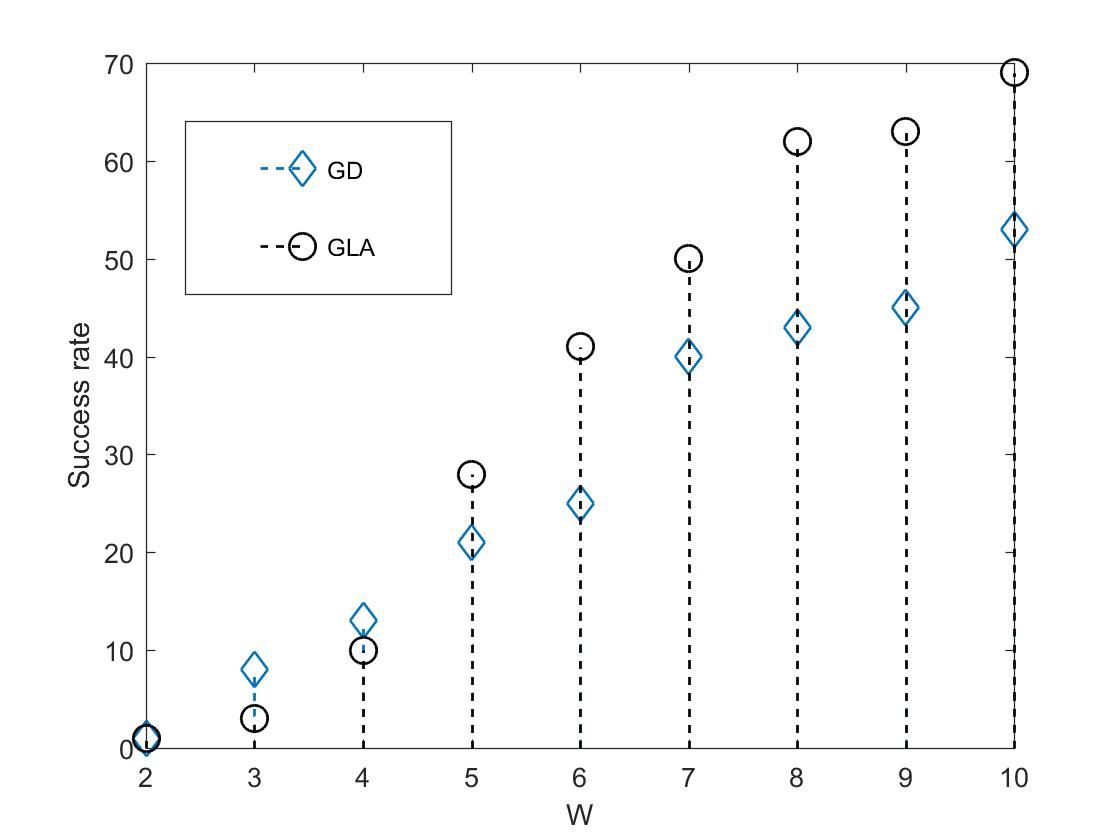}
  \caption{\label{fig:rate_of_success_stft} 
  This figure examines the empirical success rate of   a gradient algorithm (GD) that minimizes~\eqref{alg:ls}  and the Griffin-Lim algorithm (GLA) as presented in Algorithm~\ref{alg:Griffifin-Lim} for the STFT phase retrieval problem with a rectangular window. For each value of $W$, 100 experiments were conducted with $N=23$ and $L=1$ in a noise-free environment. Note that a unique solution is guaranteed according to Theorem \ref{the:per-stft-uni:2}.
  The underlying signals and the initializations were drawn from an i.i.d.\ normal distribution. A success was declared when the objective function is below $10^{-4}$. }
\end{figure}

From the algorithmic point--of--view, it is often more convenient to
formulate the problem as a minimization problem. Two 
common approaches are to minimize the intensity--based loss function
\begin{equation}
  \begin{split} \label{eq:ls}
    \mbox{min}_{z\in\mathbb{C}^N} \sum_{k=0}^{K-1}\sum_{m=0}^{M-1} \left( y[m,k]- \left\vert f_k^*D_mz\right\vert^2\right)^2,
  \end{split}
\end{equation}
or the amplitude--based loss function (see for instance \cite{fienup1982phase,wang2016solving,waldspurger2016phase})
\begin{equation}
  \begin{split} \label{eq:ls_amp}
    \mbox{min}_{z\in\mathbb{C}^N} \sum_{k=0}^{K-1}\sum_{m=0}^{M-1} \left( \sqrt{y[m,k]}- \left\vert f_k^*D_mz\right\vert\right)^2.
  \end{split}
\end{equation}
%The loss functions \eqref{eq:ls} and \eqref{eq:ls_amp} can be easily
%minimized by employing a variety of gradient algorithms\footnote{Note that
%  minimizing \eqref{eq:ls_amp} is a bit challenging as the function is
%  not smooth. However, one can employ different methods, such as
%  \cite{wang2016solving,zhang2016reshaped}.}.
 The chief difficulty  arises from the non-convexity of these loss functions. For example,
if $x$ is a real signal, then \eqref{eq:ls} is a sum of $MK$ quartic
polynomials. Hence, there is no reason to believe that a gradient
algorithm will converge to the target signal from an arbitrary
initialization.  To demonstrate this behavior, we consider an STFT phase
retrieval setup for which a unique solution is guaranteed
(see Theorem \ref{the:per-stft-uni:2}). We attempt to recover the signal by employing
two methods: a gradient descent algorithm that minimizes
\eqref{eq:ls} and the classical Griffin-Lim algorithm (see Section~\ref{sec:alternating_projection} and Algorithm
\ref{alg:Griffifin-Lim}). Both techniques were initialized from 100 different  random vectors.
As can be seen in Fig.~\ref{fig:rate_of_success_stft}, even for long windows, the algorithms do not always converge to the global minimum. Furthermore, the success rate decreases with the window's length.  
 In what follows, we present different systematic approaches to recover a signal
from its phaseless Fourier measurements and discuss their advantages
and shortcomings.

The rest of this section is organized as follows. We begin in
Section~\ref{sec:alternating_projection} by introducing the classical
algorithms which are based on alternating projections. Then, we
proceed in Section~\ref{sec:sdp} with convex programs based on semidefinite programming (SDP)
relaxations. SDPs have gained popularity in recent years as
they provide good numerical performance and theoretical guarantees. We present SDP-based algorithms for masked phase retrieval, STFT phase retrieval, minimum phase and sparse signals. In Section~\ref{sec:stft_alg}, we survey additional non-convex
algorithms with special focus on STFT phase retrieval. 
Section~\ref{sec:sparse_algorithm}  presents several   algorithms
specialized for the case of phase retrieval of sparse
signals.

\subsection{Alternating projection algorithms}  \label{sec:alternating_projection}

In their seminal work \cite{gerchberg1972practical}, Gerchberg and
Saxton considered the problem of recovering a signal from its Fourier
and temporal magnitude. They proposed an intuitive solution which iterates between two basic steps. The algorithm begins with an
arbitrary initial guess. Then, at each iteration, it imposes the known
Fourier magnitude $\vert\hat{x} \vert $ and temporal magnitude
$\vert {x} \vert $ consecutively. This process proceeds until a
stopping criterion is attained.
%{

%
%\begin{algorithm}
%  \textbf{Input:} Temporal magnitudes $\vert x\vert$ and spectral  magnitudes $\vert \hat{x}\vert $ \\ 
%  \textbf{Output:} ${x}_{est}$ - estimation of $x$ \\
%  \textbf{Initialization:} random input vector ${x}^{(0)}$, $\ell=0$\\
%  \textbf{while} halting criterion false \textbf{do}:
%  \begin{itemize}
%  \item $\ell+1 \leftarrow \ell$
%  \item Compute the Fourier transform of current estimation $\hat{x}^{(\ell)}$
%  \item Keep phase, update spectral magnitudes $\hat{z}^{(\ell)}=\vert \hat{x} \vert \operatorname{sign}(\hat{x}^{(\ell)})$
%  \item Compute ${z}^{(\ell)}$, the inverse Fourier transform of $\hat{z}^{(\ell)}$
%  \item Keep phase, update temporal magnitudes ${x}^{(\ell)}=\vert x \vert \operatorname{sign}({z}^{(\ell)})$
%  \end{itemize}
%  \textbf{end while} \\
%  \textbf{Return: } $x_{est} \leftarrow x^{(\ell)}$
%  \caption{ \label{alg:GS} Gerchberg-Saxton algorithm}
%\end{algorithm}

The basic concept of the Gerchberg-Saxton algorithm was extended by
Fienup in 1982 to a variety of phase retrieval settings
\cite{fienup1982phase,fienup1987reconstruction}. Fienup suggested to replace the temporal magnitude
constraint by other alternative constraints in the time
domain. Examples for such constraints are the knowledge of the
signal's support or few entries of the signal, non-negativity, or a known subspace in which the signal lies. Recently, it was
also suggested to incorporate sparsity priors
\cite{mukherjee2012iterative}.  These algorithms have the desired
property of \emph{error reduction}. Let $\hat{x}^{(\ell)}$ be the
Fourier transform of the estimation in the $\ell$th iteration.  Then,
it can be shown that the quantity
$E_\ell:=\sum_k\vert \vert \hat{x}[k]\vert - \vert
\hat{x}^{(\ell)}[k]\vert\vert^2 $
is monotonically non-increasing.  This class of methods is best
understood as \emph{alternating projection} algorithms \cite{elser2003phase,marchesini2007invited,pauwels2017fienup}. Namely, each
iteration consists of two consecutive projections onto sets defined by
the spectral and temporal constraints. As the first step projects onto
a non-convex set (and in some cases, the temporal projection is
non-convex as well) the iterations may not converge to the target
signal. The method is summarized in
Algorithm~\ref{alg:alternating_proj}, 
 where we use the
definition 
\begin{equation*}
  \operatorname{sign}(z[n]): =\begin{cases} \frac{z[n]}{\vert z[n] \vert},&\quad z[n]\neq 0, \\ 0,& \quad z[n]\neq 0. \end{cases}
\end{equation*}
%}

\begin{algorithm}
  \textbf{Input:} Spectral  magnitude $\vert \hat{x} \vert $  and additional temporal constraint on $x$\\ 
  \textbf{Output:} ${x}_{est}$ - estimation of $x$ \\
  \textbf{Initialization:} random input vector ${x}^{(0)}$, $\ell=0$\\
  \textbf{while} halting criterion false \textbf{do}:
  \begin{itemize}
  \item $\ell \leftarrow \ell+1$
  \item Compute the Fourier transform of current estimation $\hat{x}^{(\ell)}$
  \item Keep phase, update spectral magnitude $\hat{z}^{(\ell)}=\vert \hat{x}\vert \operatorname{sign}(\hat{x}^{(\ell)})$
  \item Compute ${z}^{(\ell)}$, the inverse Fourier transform of $\hat{z}^{(\ell)}$
  \item Impose temporal constraints on ${z}^{(\ell)}$ to obtain $x^{(\ell)}$
  \end{itemize}
  \textbf{end while} \\
  \textbf{Return: } $x_{est} \leftarrow x^{(\ell)}$

  \caption{ \label{alg:alternating_proj} General scheme of alternating projection algorithms}
\end{algorithm}

Over the years, many variants of the basic alternating projection
scheme have been suggested. A popular algorithm used for CDI
applications is the \emph{hybrid input-output} (HIO), which consists of an additional correction step in the time domain
\cite{fienup1982phase}. Specifically, the last stage of each iteration is of the form
\begin{equation*}
x^{(\ell)}[n] = \begin{cases}
z^{(\ell)}[n],&\quad  n\notin\gamma, \\ x^{(\ell-1)}[n]-\beta z^{(\ell)}[n] ,&\quad  n\in\gamma, 
\end{cases}
\end{equation*} 
where $\gamma$ is the set of indices for which $z^{(\ell)}$ violates the temporal constraint (e.g., support constraint, non-negativity) and $\beta$ is a small parameter.
While there is no proof that the HIO
converges, it tends to avoid local minima in the absence of
noise. Additionally, it is known to be sensitive to the prior
knowledge accuracy \cite{shechtman2015phase}. For additional
alternating projection schemes, we refer the interested reader to~\cite{bauschke2003hybrid, chen2007application, elser2003solution,
  luke2004relaxed, martin2012noise, rodriguez2013oversampling}.

Griffin and Lim proposed a modification of Algorithm \ref{alg:alternating_proj} 
specialized for STFT phase retrieval
\cite{griffin1984signal}. In this approach, the last step at each
iteration harnesses the knowledge of the STFT window to update the signal estimation. The Griffin-Lim heuristic is summarized in
Algorithm~\ref{alg:Griffifin-Lim}.

\begin{algorithm}
  \textbf{Input:} STFT  magnitude $ \vert \hat{x}_d[m,k] \vert $ \\ 
  \textbf{Output:} ${x}_{est}$ - estimation of $x$ \\
  \textbf{Initialization:} random input vector ${x}^{(0)}$, $\ell=0$ \\
  \textbf{while} halting criterion false \textbf{do}:
  \begin{itemize}
  \item $\ell \leftarrow \ell+1$
  \item Compute the STFT of current estimation $\hat{x}_d^{(\ell)}$
  \item Keep phase, update STFT magnitudes $\hat{z}^{(\ell)}=\vert \hat{x}_d \vert\operatorname{sign}(\hat{x}_d^{(\ell)})$
  \item For each fixed $m$, compute $z^{(\ell)}_m$, the inverse Fourier transform of $\hat{z}^{(\ell)}$ 
  \item Update signal estimate  ${x}^{(\ell)}[n] = \frac{\sum_m z^{(\ell)}_m[n]\,\overline{d[mL-n]}}{\sum_m\vert d[mL-n]\vert^2}$
  \end{itemize}
  \textbf{end while} \\
  \textbf{Return: } $x_{est} \leftarrow {x}^{(\ell)}$

  \caption{ \label{alg:Griffifin-Lim} Griffin-Lim algorithm}
\end{algorithm}

\subsection{Semidefinite relaxation algorithms} \label{sec:sdp}
In recent years,  algorithms based on convex relaxation
techniques have attracted considerable attention \cite{candes2015phase,waldspurger2015phase}. These methods are
based on the insight that while the feasibility problem
\eqref{eq:pr_feas} is quadratic with respect to $x$, it is linear
in  the matrix $xx^*$. This leads to a natural convex SDP relaxation that can be solved in 
polynomial time using standard solvers like CVX
\cite{grant2008cvx}. In many cases, these relaxations achieve
excellent numerical performance followed by theoretical guarantees. However, the SDP relaxation optimizes over $N^2$ variables and therefore its computational complexity is quite high.

SDP relaxation techniques begin by reformulating the measurement model
\eqref{eq:PR_vector} as a linear function of the Hermitian rank--one
matrix $X:=xx^*$:
\begin{equation*}
  \begin{split}
    y[m,k] &= (f_k^*D_mx)^*(f_k^*D_mx) = x^*D_m^*f_kf_k^*D_mx = \operatorname{trace}(D_m^*f_kf_k^*D_mX). 
  \end{split}
\end{equation*} 
Consequently, the problem of recovering  $x$ from $y$ can be posed
as the feasibility problem of finding a rank--one Hermitian matrix
which is consistent with the measurements:
\begin{equation} \label{eq:feas_rank}
  \begin{split}
    \mbox{find}\quad X\in\mathcal{H}^N \quad \mbox{subject to} \quad& X\succeq 0,\quad \operatorname{rank}(X)=1, \\
    &y[m,k] = \operatorname{trace}(D_m^*f_kf_k^*D_mX), \\ &k=0,\dots,K-1,\quad m=0,\dots,M-1, 
  \end{split} 
\end{equation} 
where $\mathcal{H}^N$ is the set of all $N\times N$ Hermitian
matrices.  If there exists a matrix $X$ satisfying all the constraints
of \eqref{eq:feas_rank}, then it determines $x$ up to global phase.  The
feasibility problem \eqref{eq:feas_rank} is non-convex due to the rank
constraint. A convex relaxation may be obtained by omitting the rank
constraint leading to the SDP~\cite{candes2015phase, goemans1995improved, shechtman2011sparsity,waldspurger2015phase}:
\begin{equation} \label{eq:sdp_relax}
  \begin{split}
    \mbox{find}\quad X\in\mathcal{H}^N \quad \mbox{subject to} \quad& X\succeq 0, \\
    &y[m,k] = \operatorname{trace}(D_m^*f_kf_k^*D_mX), \\ &k=0,\dots,K-1,\quad m=0,\dots,M-1.
  \end{split} 
\end{equation}

If the solution of \eqref{eq:sdp_relax} happens to be of rank one,
then it determines $x$ up to global phase. In practice, it is useful to
promote a low rank solution by minimizing an objective function over
the constraints of \eqref{eq:sdp_relax}.  A typical choice  is the trace function, which is the convex hull of the rank function 
for Hermitian matrices. The resulting SDP relaxation algorithm is summarized in
Algorithm~\ref{alg:sdp}.
% We do not dive here into the technical details of how to how that
% the SDP relaxation \eqref{eq:sdp_relax} attains the solution of
% \eqref{eq:feas_rank}. We just mention briefly that it requires
% showing the existence of a special matrix, frequently called
% \emph{dual certificate}. The existence of the dual certificate
% guarantees that the KKT conditions are met, see for instance
% \cite{vandenberghe1996semidefinite}.

\begin{algorithm}
  \textbf{Input:} Fourier  magnitudes $y[m,k]$ as given in \eqref{eq:PR}
  and the masks $D_m,\ m=0,\dots,M-1$ \\
  \textbf{Output:} ${x}_{est}$ - estimation of $x$ \\
  \textbf{Solve}: \\
  \begin{equation*} 
    \begin{split}
      \min_{x\in\mathcal{H}^N }\mbox{trace}(X) \quad \mbox{subject to} \quad& X\succeq 0, \\[-6pt]
      &y[m,k] = \operatorname{trace}(D_m^*f_kf_k^*D_mX), \\ &k=0,\dots,K-1,\quad m=0,\dots,M-1.
    \end{split} 
  \end{equation*} 
  \textbf{Return :} $x_{est}$ - the best rank--one approximation of the SDP's solution.
  \caption{ \label{alg:sdp} SDP relaxation for phase retrieval with masks}
\end{algorithm}

The SDP relaxation for the classical phase retrieval problem
(i.e., $M=1$ and $D_0=I_N$) was investigated in \cite{haung2016}. It
was shown that SDP relaxation achieves the optimal cost function
value of (\ref{eq:ls}). 
However, recall that in general the classical phase retrieval problem  does not admit a unique solution.  Minimum phase signals are an exception as explained in  Section \ref{sec:minim-phase-sign}. Let  $a$ be the autocorrelation sequence of the estimated signal from Algorithm \ref{alg:sdp}.
 If $x$ is minimum phase, then it can be recovered by the following program:
\begin{equation} \label{eq:sdp_MP}
  \begin{split}
    \operatorname{max}_{X\in\mathcal{H}^N} \quad X[0,0]\quad  \mbox{subject to} \quad& X\succeq0,\quad \operatorname{trace}(\Theta_kX)=a[k], \\
    & k=0,\dots,N-1,
  \end{split} 
\end{equation} 
where $\Theta_k$ is a Toeplitz matrix with ones in the $k$th diagonal
and zero otherwise.  The solution of \eqref{eq:sdp_MP}, $X_{MP}$, is guaranteed to be rank one so that $X_{MP}=xx^*$. 
See Section \ref{sec:minim-phase-sign} for a different algorithm to recover minimum phase signals.

An SDP relaxation for deterministic  masks was investigated in
\cite{jaganathan2015phase_mask}, where the authors consider
two types of masks. Here, we consider the two masks, $d_1$ and $d_2,$ given in
\eqref{eq:fixed_masks}.  Let $D_1$ and $D_2$ be the diagonal matrices
associated with $d_1$ and $d_2$, respectively, and  assume that
each measurement is contaminated by bounded noise $\varepsilon$. Then, it was
suggested to estimate the signal by solving the following convex
program 
\begin{equation} \label{eq:sdp_relax_masked}
  \begin{split}
    \operatorname{min}_{X\in\mathcal{H}^N} \quad \operatorname{trace}(X)\quad  \mbox{subject to} \quad& X\succeq 0, \\
    &\vert y[m,k] - \operatorname{trace}(D_m^*f_kf_k^*D_mX)\vert \leq\varepsilon, \\ &\thinspace k=0,\dots,2N-1,\quad m=0,1.
  \end{split} 
\end{equation} 
 This program achieves
 stable recovery  in the sense that the
recovery error is proportional to the noise level and reduces to zero
in the noise-free case. Note however that in the presence of noise the solution is not likely to be  rank one.

\begin{thm}
  {\upshape\cite{jaganathan2015phase_mask}} Consider a signal
  $x\in\mathbb{C}^N$ satisfying $\Vert x\Vert_1\leq \beta$ and
  $\vert x[0]\vert\geq \gamma>0$. Suppose that the measurements are
  taken with the diagonal matrices $D_1$ and $D_2$
  {\upshape(}masks{\upshape)} associated with $d_1$ and $d_2$ given in
  \eqref{eq:fixed_masks}. Then, the solution $\widetilde X$ of the
  convex program \eqref{eq:sdp_relax_masked} obeys
  \begin{equation*}
    \Vert \widetilde X-xx^*\Vert_2\leq C(\beta,\gamma)\varepsilon
  \end{equation*}
  for some numerical constant $C(\beta,\gamma)$.
\end{thm}

Phase retrieval from STFT measurements using SDP was
considered in \cite{jaganathan2016stft}. Here, SDP relaxation in
the noiseless case takes on the form 
\begin{equation} \label{eq:sdp_relax_stft}
  \begin{split}
    \operatorname{min}_{X\in\mathcal{H}^N} \quad \operatorname{trace}(X)\quad  \mbox{subject to} \quad& X\succeq 0, \\
    & y[m,k]= \operatorname{trace}(D_m^*f_kf_k^*D_mX), \\ &\thinspace k=0,\dots,K-1,\quad m=0,\dots,M-1, 
  \end{split} 
\end{equation} 
where $M=\lceil N/L \rceil$ is the number of STFT windows and $\tilde{N}=N$ (see \eqref{eq:PR_STFT}). In \cite{jaganathan2016stft}, it was
proven that \eqref{eq:sdp_relax_stft} recovers the signal
exactly under the following conditions.

\begin{thm}
  \label{th:stft_sdp}
  {\upshape\cite{jaganathan2016stft}} The convex program
  \eqref{eq:sdp_relax_stft} has a unique feasible matrix $X=xx^*$ for
  almost all non-vanishing signals $x$ if:
  \begin{itemize}
  \item  $d[n]\neq 0$ for $n=0,\dots,W-1$,
  \item $2L \leq W\leq N/2 $,
  \item $4L\leq K\leq N$,
  \item $x[n]$ is known apriori for $0\leq n\leq  \left\lfloor\frac{L}{2} \right\rfloor$,
  \end{itemize} 
where $W$ is the window's length.
\end{thm}
Note that for $L=1$ no prior knowledge on the entries of $x$ is required.

\begin{figure} 
  \centering
  \includegraphics[scale=0.6]{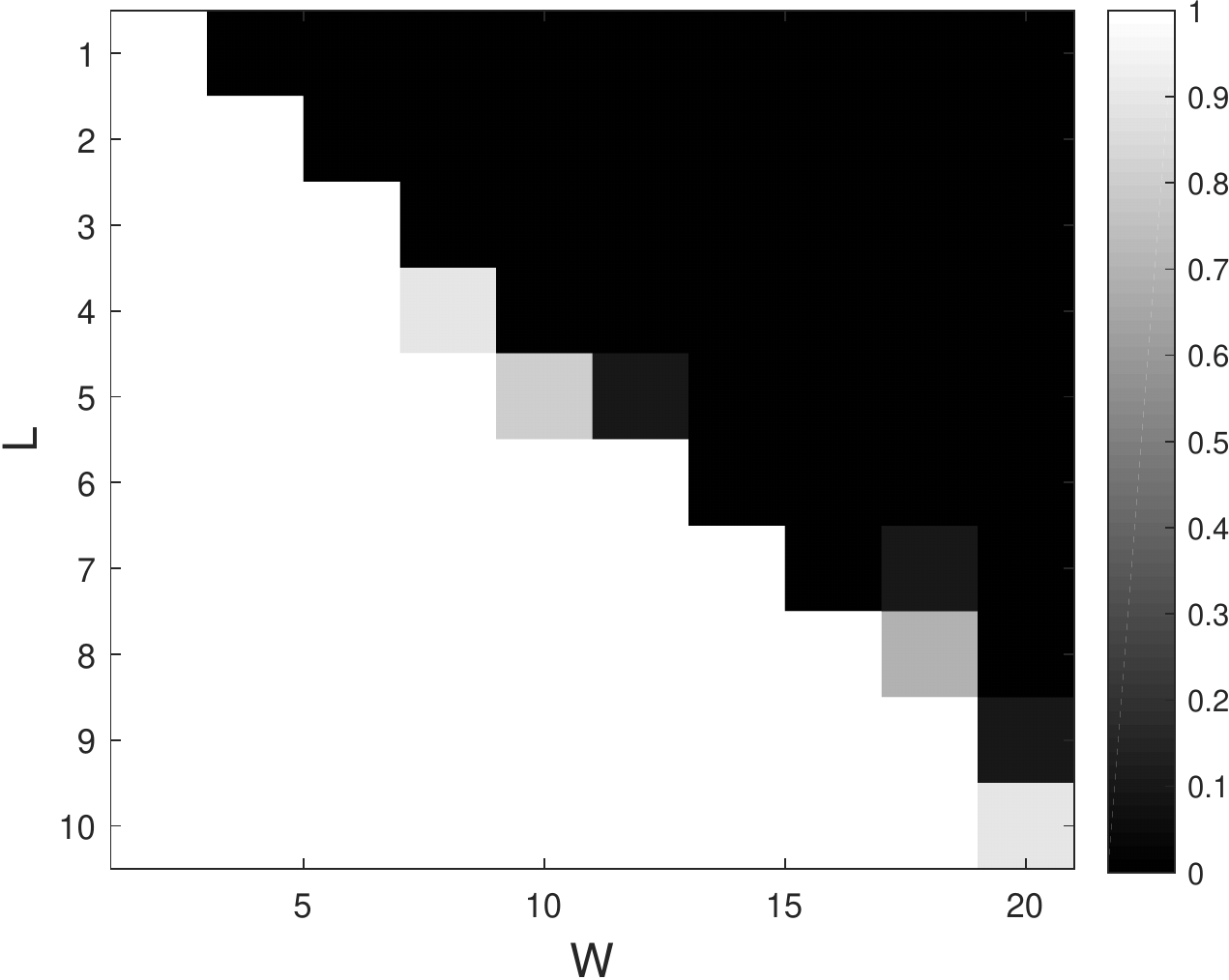}
  \caption{\label{fig:sdp_stft} The empirical success rate of the SDP
    relaxation for STFT phase retrieval \eqref{eq:sdp_relax_stft} with
    a rectangular window of length $W$, i.e.,\ $d[n]=1$ for
    $n=0,\dots,W-1$. For each pair $(W,L)$, 100 complex signals of
    length $N=40$ were drawn from an i.i.d.\ normal distribution. The
    figure presents the empirical success rate. An experiment was
    declared as a success if the recovery error is below $10^{-4}$.  }
\end{figure}

An interesting implication of this thm is that recovery remains exact even if we merely have 
 access to the low frequencies of the data. This property is
called \emph{super-resolution} and will be discussed in more detail in
the context of sparse signals.  Numerically, the performance of
\eqref{eq:sdp_relax_stft} is better than 
Theorem~\ref{th:stft_sdp} suggests. Specifically, it seems that for $W\geq 2L$,
\eqref{eq:sdp_relax_stft} recovers $xx^*$ exactly without any prior knowledge on the entries of $x$, as demonstrated in 
Fig.~\ref{fig:sdp_stft} (a similar example is given in \cite{jaganathan2016stft}). Additionally, the program is stable in the
presence of noise.

\subsection{Additional non-convex algorithms} \label{sec:stft_alg}

In this section, we present additional non-convex algorithms for phase retrieval with special focus on STFT phase retrieval. A naive way to estimate the signal from its phaseless measurements is by minimizing the 
non-convex loss functions \eqref{eq:ls} or \eqref{eq:ls_amp}  by employing a gradient descent
scheme. However, as demonstrated in Fig.~\ref{fig:rate_of_success_stft},  this algorithm is likely to converge to a local minimum due
to the non-convexity of the loss functions. Hence, the key is to
introduce an efficient method to initialize the non-convex algorithm
sufficiently close to the global minimum.   

A recent paper \cite{bendory2016non} suggests  an initialization technique for STFT phase retrieval, which we now describe. Consider
the one-dimensional Fourier transform of the data with respect to the
frequency variable (see \eqref{eq:PR_STFT}), given by
\begin{equation*}
  \tilde{y}[m,\ell]
  =\sum_{n=0}^{N-1}{x}[n]\overline{x[n+\ell]}d[mL-n]\overline{d[mL-n-\ell]},
\end{equation*}
where $\tilde{N}=N$ and both the signal and the window are assumed to be periodic.
For fixed $\ell$, we obtain the linear system of equations
\begin{equation} \label{eq:Gl}
  \tilde{y}_\ell=G_\ell x_\ell,
\end{equation}
where $\tilde{y}_\ell=\{\tilde{y}[m,\ell]\}_{m=0}^{\lceil N/L\rceil -1}$, $x_\ell\in\mathbb{C}^{N}$ is the
$\ell$th diagonal of the matrix $xx^*$  and 
 $G_\ell\in\mathbb{C}^{\lceil N/L\rceil \times N}$ is the matrix with $(m,n)$th entry given by $d[mL-n]\overline{d[mL-n-\ell]}$. For $L=1$, $G_\ell$ is a circulant matrix. Clearly, recovering $x_\ell$
for all $\ell$ is equivalent to recovering $xx^*$. Hence, the ability to
estimate $x$ depends on the properties of the window which determines $G_\ell$. To
make this statement precise, we use the following definition.

\begin{defn}
  \label{def:admissible_win}A window ${d}$ is called an
  \emph{admissible window of length $W$} if for all
  $\ell=-(W-1),\dots,(W-1)$ the associated circulant matrices
  ${G}_{\ell}$  in \eqref{eq:Gl} are invertible.
\end{defn}
An example of an admissible window is a rectangular window with
$W\leq N/2$ and $N$ a prime number.  If the STFT window is
sufficiently long and admissible, then the STFT phase retrieval has a
closed-form solution. This solution can be obtained by the principal
eigenvector of a matrix, constructed as the solution of a
least-squares problem according to \eqref{eq:Gl}. This algorithm is
summarized in Algorithm~\ref{alg:ls}.

\begin{thm}
  {\upshape\cite{bendory2016non}} Let $L=1$ and suppose that $d$ is an
  admissible window of length
  $W\geq\left \lceil\frac{N+1}{2}\right\rceil$ {\upshape(}see
  Definition~\ref{def:admissible_win}{\upshape)}.  Then,
  Algorithm~\ref{alg:ls} recovers any complex signal up to global phase.
\end{thm}

\begin{algorithm}
  \textbf{Input:} The STFT magnitude $y[m,k]$ as given in \eqref{eq:PR_STFT} with $\tilde{N}=N$\\
  \textbf{Output:} ${x}_{est}$ - estimation of $x$ 
  \begin{enumerate}
  \item Compute $\tilde{y}\left[m,\ell\right]$, the one-dimensional DFT of $y[m,k]$ with respect to the second variable.

  \item Construct a matrix ${X}_{0}$ such that 
    \[
    \operatorname{diag}\left({X}_{0},\ell\right)=\begin{cases}
      {G}_{\ell}^{\dagger}\tilde{y}_{\ell}, & \ell=-\left(W-1\right),\cdots,\left(W-1\right),\\
      0, & \mbox{otherwise,}
    \end{cases}
    \]
    where ${G}_{\ell}\in\mathbb{R}^{N\times N}$ and $\tilde{y}_{\ell}$ are given  in \eqref{eq:Gl}.  
  \end{enumerate}
  \textbf{Return: }
  \[ {x}_{est}=\sqrt{\sum_{n\in
      P}({G}_{0}^{\dagger}{y}_{0})[n]}{x}_{p},
  \]
  where $P:=\{ n\thinspace:\thinspace({G}_{0}^{\dagger}{y}_{0})[n]>0\} $ and ${x}_{p}$ is the principle (unit-norm) eigenvector of
  $X_0$.
  \protect\caption{\label{alg:ls}Least-squares algorithm for STFT phase retrieval with $L=1$}
\end{algorithm}

In many cases, the window is shorter than
$ \left\lceil\frac{N+1}{2}\right\rceil$. However, the same technique
can be applied to initialize a refinement process, such as a gradient
method or the Griffin-Lim algorithm (GLA). In this case, the distance
between the initial vector (the output of Algorithm~\ref{alg:ls}) and
the target signal can be estimated as follows.

\begin{thm}
  {\upshape\cite{bendory2016non}} Suppose that $L=1$, $\|x\|_2=1$, $d$
  is an admissible window of length $W\geq 2$ and that
  $\|{x}\|_{\infty}\leq\sqrt{{B}/{N}}$ for some
  $0<B\leq{N}/{\left(2N-4W+2\right)}$. Then, the output $x_0$ of 
   Algorithm~\ref{alg:ls} satisfies
  \[
  \min_{\phi\in[0,2\pi)}\|x-x_0e^{j\phi}\|_2^2\leq 1-\sqrt{1-2B \, \frac{N-2W+1}{N}}.
  \]
\end{thm}

\begin{figure}
  \begin{minipage}[t]{.5\columnwidth}%
    % \subfloat[Initialization with linear interpolation]
    \includegraphics[scale=0.33]{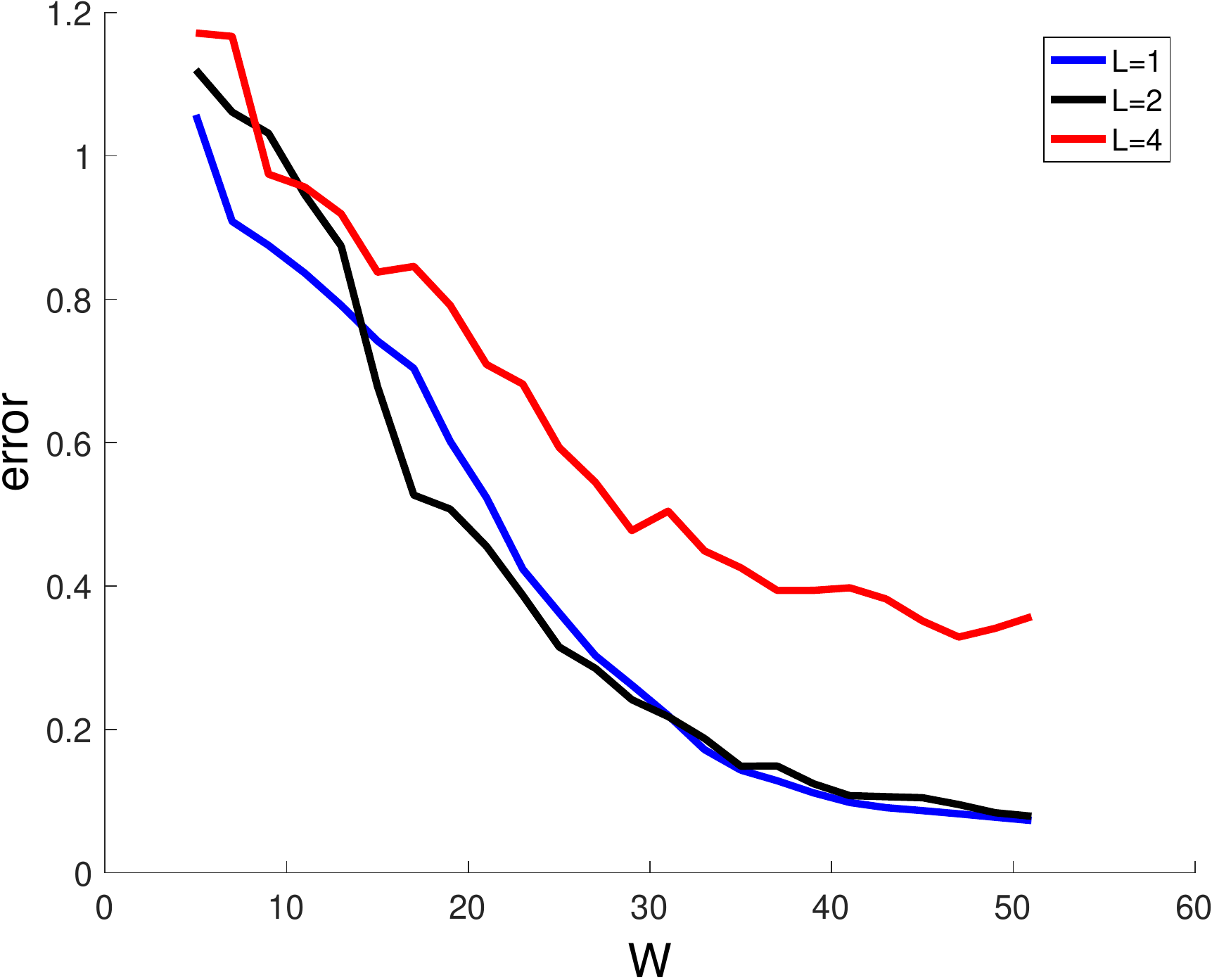}
    % {
    % }%
  \end{minipage}%
  \begin{minipage}[t]{.33\columnwidth}%
    % \subfloat[L=2]{\includegraphics[scale=0.4]{gd_gla_L2}

    % }%
    \includegraphics[scale=0.3]{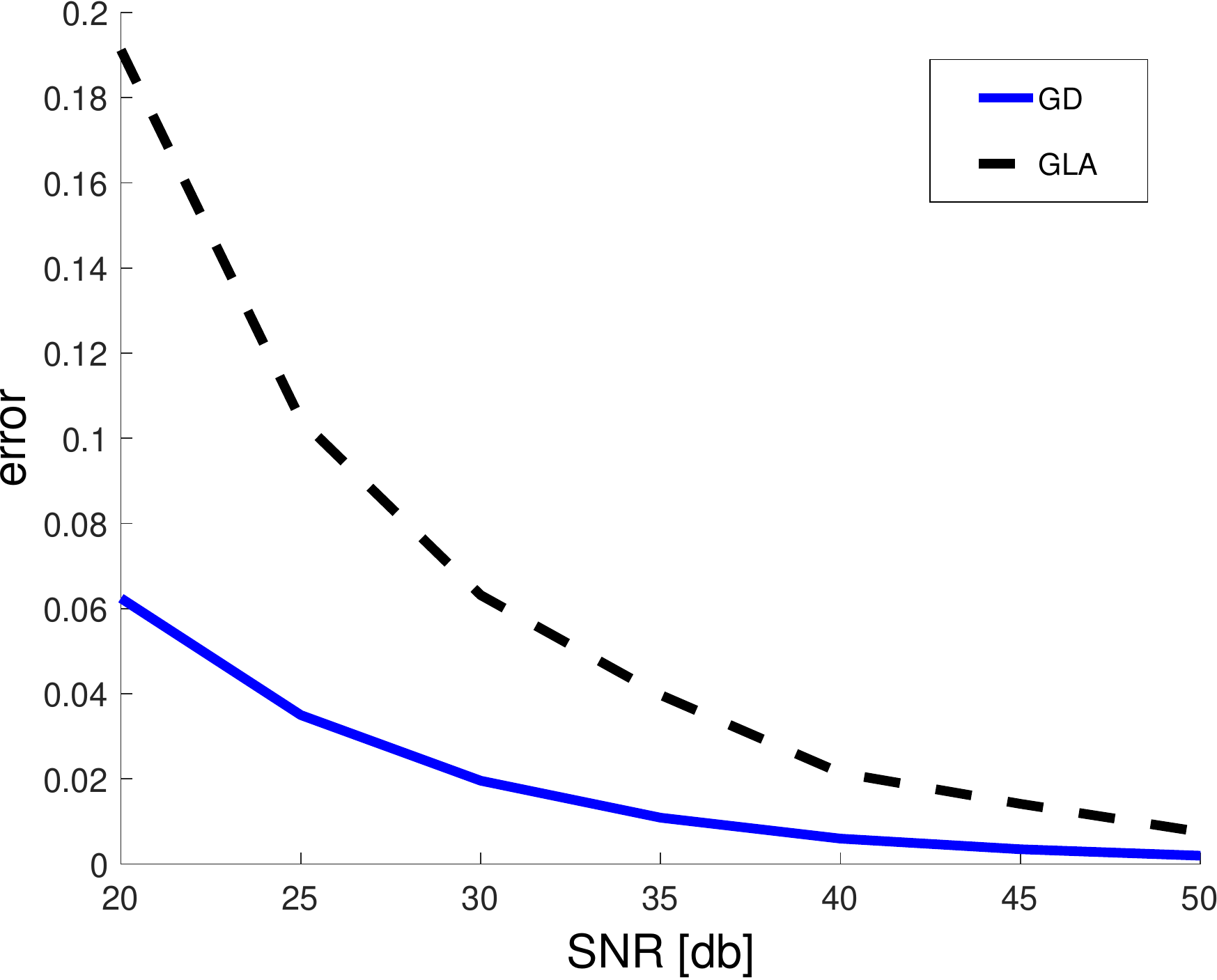}
  \end{minipage}

  \protect\caption{\label{fig:init} \textbf{(left)} Average error
    (over 50 experiments) of the initialization method of
    Algorithm~\ref{alg:ls} as a function of $W$ and $L$. The
    experiments were conducted on a signal of length $N=101$ with a
    Gaussian window $d[n]=e^{-{n^{2}}/{2\sigma^{2}}}$. The window
    length was set to $W=3\sigma$.  \textbf{(right)} Average normalized 
    recovery error (over 20 experiments) of the gradient descent (GD) and
    Griffin-Lim algorithm (GLA) in the presence of normal i.i.d.\
    noise. Both algorithms were initialized by
    Algorithm~\ref{alg:ls}. The experiments were conducted on signals
    of length $N=53$ with a rectangular window of length $W=19$ and
    $L=2$. The figures are taken from \cite{bendory2016non}. }
\end{figure}

For $L>1$, it is harder to obtain a reliable estimation of the
diagonals of $xx^*$. Nevertheless, a simple heuristic is proposed in
\cite{bendory2016non} based on the smoothing properties of typical
STFT windows. %The algorithm also works when the high-frequencies of
%the acquired data are annihilated.
%  For details, the reader iscreferred to \cite{bendory2016non}.  
Fig.~\ref{fig:init} shows
experiments corroborating the effectiveness of this initialization
approach for $L>1$.

We have seen that under some conditions, the STFT phaseless
measurements provide partial information on the matrix $xx^*$. In
some cases, the main diagonal of $xx^*$, or equivalently the temporal
magnitude of $x$, are also measured. Therefore, if the signal is
non-vanishing, then all entries of the matrix $xx^*$ can be normalized to have
unit modulus.  This in turn implies that the STFT phase retrieval
problem is equivalent to estimating the missing entries of a rank--one
matrix with unit modulus entries (i.e.,\ phases). This problem is known
as \emph{phase synchronization}.
In recent years, several algorithms for phase synchronization were  suggested and
analyzed, among them eigenvector-based methods, SDP relaxations,
projected power methods and approximate message passing algorithms
\cite{bandeira2014tightness, bandeira2015non, boumal2016nonconvex,
  chen2016projected, perry2016message, singer2011angular}.  Recent
papers \cite{iwen2016phase, iwen2016fast} adopted this approach and
suggested spectral and greedy algorithms for STFT phase
retrieval. These methods are accompanied by stability guarantees and
can be modified for phase retrieval using masks. The main shortcoming
of this approach is that it relies on a good estimation of the
temporal magnitudes which may not always be available.

Another interesting approach has been recently proposed in
\cite{pfander2016robust}. This paper suggests a multi-stage algorithm
based on spectral
clustering and angular synchronization. It is shown that the algorithm achieves stable
estimation (and exactness in the noise-free setting) with only
$O(N\log N)$ phaseless STFT measurements. Nevertheless, the algorithm
builds upon random STFT windows of length $N$ while most applications
use shorter windows.

\subsection{Algorithms for sparse signals} \label{sec:sparse_algorithm}

In this section, we assume that the signal is sparse with a sparsity level defined as
\begin{equation*}
  s = \{\#n \thinspace : \thinspace x[n] \neq 0\}.
\end{equation*}
In this case, the basic phase retrieval problem \eqref{eq:ls} can be
modified to the constrained least-squares problem
\begin{equation}
  \begin{split} \label{eq:ls_sparse}
    \operatorname{min}_{z\in\mathbb{C}^N} \sum_{k=0}^{K-1}\sum_{n=0}^{M-1} \left( y[m,k]- \left\vert f_k^*D_mz\right\vert^2\right)^2 \quad\mbox{subject to}\quad \|z\|_0\leq s,
  \end{split}
\end{equation}
where we use $\|\cdot\|_0$ as the standard $\ell_0$ pseudo-norm
counting the non-zero entries of a signal.

Many phase retrieval algorithms for sparse signals are modifications
of known algorithms for the non-sparse case. For instance, gradient algorithms where modified to take into account the sparsity structure. The underlying idea of these algorithms is to add a thresholding step at each iteration. 
Theoretical analysis of these algorithms for phase retrieval with random sensing vectors is considered in \cite{cai2016optimal,wang2016sparse}. 
A similar modification for the HIO
algorithm was proposed in 
\cite{mukherjee2012iterative}.  Modifications of SDP relaxation
methods for phase retrieval with random sensing vectors were
considered in
\cite{li2013sparse, ohlsson2011compressive, oymak2015simultaneously}.
Here, the core idea is to incorporate a sparse-promoting regularizer in the
objective function. However, this technique cannot be adapted directly
to Fourier phase retrieval because of the trivial ambiguities of
translation and conjugate reflection; see a detailed explanation in
\cite{jaganathan2015phase}. To overcome this barrier, a Two-Stage 
Sparse Phase-Retrieval (TSPR) algorithm was proposed in
\cite{jaganathan2013sparse}. The first stage of the algorithm involves
estimating the support of the signal directly from the support of its
autocorrelation. This problem is equivalent to the \emph{turnpike
  problem} of estimating a set of integers from their pairwise
distances \cite{skiena1990reconstructing}. Once the support is known,
the second stage involves solving an SDP to estimate the missing
amplitudes. It was proven that TSPR recovers signals exactly in the
noiseless case as long as the sparsity level is less than
$O(N^{1/2})$. In the noisy setting, recovery is robust for sparsity
level lower than $O(N^{1/4})$. A different SDP-based approach was
suggested in \cite{shechtman2011sparsity}. This method proposes to
promote a sparse solution by the log-det heuristic \cite{fazel2003log}
and an $\ell_1-\ell_2$ constraint on the matrix $xx^*$.

An alternative class of algorithms that has been proven to be highly
effective for sparse signals is the class of greedy algorithms, see
for instance \cite{beck2013sparsity, mallat1993matching}. For phase
retrieval tasks, a greedy optimization algorithm called GESPAR (GrEedy
Sparse PhAse Retrieval) is proposed in
\cite{shechtman2014gespar}. The algorithm was applied for a variety of
optical tasks, such as CDI and phase retrieval through waveguide
arrays \cite{shechtman2013efficient, shechtman2013sparsity,
  sidorenko2013sparsity,sidorenko2015sparsity}.  GESPAR is a local search algorithm, based
on iteratively updating the signal support.  Specifically, two
elements are being updated at each iteration by swapping. Then, a
non-convex objective function that takes the support into account is
minimized by a damped Gauss-Newton method.  The swap is carried out
between the support element corresponds to the smallest entry (in
absolute value) and the off-support element with maximal gradient
value of the objective function. A modification of GESPAR for STFT
phase retrieval was presented in \cite{eldar2015sparse}.  A schematic
outline of GESPAR is given in Algorithm~\ref{alg:GESPAR}; for more details,
see \cite{shechtman2014gespar}.

\begin{algorithm}
  \textbf{Input:} Fourier magnitude $y$ as in \eqref{eq:PR} and sparsity level $s$  \\
  \textbf{Output:} ${x}_{est}$ - estimation of $x$ \\
  \textbf{Initialization:}
  \begin{itemize}
  \item Generate a random support set $S^{(0)}$ of size $s$
  \item Employ a damped Gauss-Newton method with support $S^{(0)}$ and obtain an initial estimation $x^{(0)}$
  \item Set $\ell=0$
  \end{itemize}
  \textbf{while} halting criterion false \textbf{do}:
  \begin{itemize}
  \item $\ell \leftarrow \ell+1$
  \item Update support by swapping two entries, one in $S^{(\ell-1)} $
    and one in the complementary set
  \item Minimize a non-convex objective with the given support $S^{(\ell)}$ using  the damped Gauss-Newton method to obtain $ x^{(\ell)}$
  \end{itemize}
  \textbf{end while} \\
  \textbf{Return:} $x_{est} \leftarrow x^{(\ell)}$
  \caption{ \label{alg:GESPAR} A schematic outline of GESPAR algorithm; for details see 
    \cite{shechtman2014gespar}}
\end{algorithm}

In practice, many optical measurement processes blur the fine details
of the acquired data and act as low-pass filters. In these cases, one
aims at estimating the signal from its low-resolution Fourier
magnitudes. This problem combines two classical problems: phase
retrieval and super-resolution. In recent years, 
super-resolution for sparse signals has been investigated thoroughly \cite{azais2015spike,bendory2017robust,bendory2015super, bendory2016robust, candes2014towards,   duval2015exact}.  In
Theorem~\ref{th:stft_sdp}, we have seen that the SDP
\eqref{eq:sdp_relax_stft} can recover a signal from its low-resolution
STFT magnitude.  The  problem
of recovering a signal from its low-resolution phaseless measurements
using masks was considered in \cite{jaganathan2016phaseless,salehi2017multiple}. It was 
proven that exact recovery may be obtained by only few\footnote{Specifically, several combinations of masks are suggested. Each combination consists of three to five deterministic masks. } carefully designed masks if 
the underlying signal is sparse and its support is not clustered (this
requirement is also known as the separation condition). An extension to 
the continuous setup was suggested in \cite{cho2016phaseless}. A
combinatorial algorithm for recovering a signal from its
low-resolution Fourier magnitude was suggested in
\cite{chen2014algorithm}. The algorithm recovers an $s$-sparse signal
exactly from $2s^2-2s+2$ low-pass magnitudes. Nevertheless, this
algorithm is unstable in the presence of noise due to error
propagation.

%%% Local Variables:
%%% mode: latex
%%% TeX-master: "draft4"
%%% End:

\section{Conclusion} \label{sec:conclusion}

In this chapter, we studied the problem of Fourier phase retrieval. We focused on the question of uniqueness, presented the main algorithmic approaches and discussed their properties. 
 To conclude the chapter, we  outline  several fundamental gaps in the theory of Fourier phase retrieval.

 Although many different methods have been proposed and
  analyzed in the last decade for Fourier phase retrieval,  alternating projection  algorithms maintained their popularity. Nevertheless, the theoretical understanding of these
  algorithms is limited. 
   Another fundamental open question regards multidimensional phase retrieval. While almost all multidimensional signals are determined uniquely by their Fourier magnitude,  there is no method that provably recovers the signal.

%Another important question refers to the design and analysis of efficient algorithms for the combined problem of phase retrieval and super-resolution.
 In many applications in optics, the measurement process acts as
  a low-pass filter. Hence, a practical algorithm should recover the
  missing phases (phase retrieval) and resolve the fine details of the
  data (super-resolution).  In this chapter, we surveyed several works
  dealing with the combined problem. Nonetheless, the current
  approaches are based on inefficient SDP programs
  \cite{cho2016phaseless, jaganathan2016stft,jaganathan2016phaseless,salehi2017multiple} or lack
  theoretical analysis \cite{bendory2016non,
    shechtman2011sparsity}.
Additionally, even if all frequencies are available, it is still not clear what is the maximal sparsity that enables efficient and stable recovery of a  signal from its Fourier magnitude.  
% As of today, there is no efficient algorithm that solve the phase retrieval and
 % super-resolution problems simultaneously with theoretical analysis.

%In several optical applications, the problem of  STFT phase retrieval arises. However, there is still no provable efficient method for the recovery with short windows.
 In ultra short laser pulse characterization, it is  common to use the FROG methods that were introduced in Section \ref{sec:FROG}.  It is 
  interesting to understand  the minimal number of measurements
  which can guarantee uniqueness for FROG-type methods. Additionally,
  a variety of algorithms are applied  to estimate signals
  from FROG-type measurements; a theoretical understanding of these
  algorithms is required.
 
%%% Local Variables:
%%% mode: latex
%%% TeX-master: "draft1"
%%% End:

%~~~~~~~~~~~~~~~~~~~~~~~~~~~~~~~~~~~~~~~~~~~~~~~~~~~~~~~~~~~~~%

 \bibliographystyle{plain}
  
 \bibliography{ref}

\end{document}